\documentclass[manuscript]{acmart}
\usepackage{subcaption}
\usepackage[ruled,vlined,linesnumbered]{algorithm2e}
\usepackage{xspace}
\usepackage{amsmath,amsfonts}
\usepackage{hyperref}
\usepackage{hyperxmp}
  
\usepackage{amssymb}
\usepackage{algorithmic}
\usepackage{graphicx}
\usepackage{textcomp}
\usepackage{xcolor}
\usepackage{booktabs}
\usepackage{multirow}
\usepackage{caption}
\usepackage{dcolumn}
\usepackage{wrapfig}
\definecolor{mygray}{RGB}{247,247,247}
\usepackage{enumitem}
\usepackage{listings}
\usepackage{xcolor}
\usepackage[capitalize]{cleveref}
\crefname{section}{Sec.}{Secs.}
\crefname{section}{Section}{Sections}
\crefname{table}{Table}{Tables}
\crefname{table}{Tab.}{Tabs.}
\usepackage{tcolorbox}
\newcommand{\mybox}[1]{%
  \begin{tcolorbox}[colback=mygray,colframe=black,lowerbox=invisible,savelowerto=\jobname_ex.tex]
    \emph{#1}
  \end{tcolorbox}
}
\definecolor{codegreen}{rgb}{0,0.6,0}
\definecolor{codegray}{rgb}{0.5,0.5,0.5}
\definecolor{codepurple}{rgb}{0.58,0,0.82}
\definecolor{backcolour}{rgb}{0.95,0.95,0.92}
\newcommand{\dataset}{ULT\xspace}
\newcommand{\lkdataset}{PLT\xspace}
\lstdefinestyle{mystyle}{
  backgroundcolor=\color{backcolour},   commentstyle=\color{codegreen},
  keywordstyle=\color{magenta},
  numberstyle=\tiny\color{codegray},
  stringstyle=\color{codepurple},
  basicstyle=\ttfamily\footnotesize,
  breakatwhitespace=false,         
  breaklines=true,                 
  captionpos=b,                    
  keepspaces=true,                 
  numbers=left,                    
  numbersep=5pt,                  
  showspaces=false,                
  showstringspaces=false,
  showtabs=false,                  
  tabsize=2
}

\begin{document}

\title{Benchmarking LLMs for Unit Test Generation from Real-World Functions}

\author{Dong HUANG}
\email{dhuang@nus.edu.sg}
\affiliation{%
  \institution{National University of Singapore}
  \country{Singapore}
}

\author{Jie M. Zhang}
\email{jie.zhang@kcl.ac.uk}
\affiliation{%
  \institution{King's College London}
  \country{London, UK}
}
\author{Mark Harman}
\email{mark.harman@ucl.ac.uk}
\affiliation{%
  \institution{University College London}
  \country{London, UK}
}
\author{Qianru Zhang}
\email{qz348@cam.ac.uk}
\affiliation{%
  \institution{The University of Cambridge}
  \country{Cambridge, UK}
}
\author{Mingzhe Du}
\authornote{Corresponding author.}
\email{mingzhe@nus.edu.sg}
\affiliation{%
  \institution{National University of Singapore}
  \country{Singapore}
}
\author{See-Kiong Ng}
\email{seekiong@nus.edu.sg}
\affiliation{%
  \institution{National University of Singapore}
  \country{Singapore}
}

\begin{abstract}
Recently, large language models (LLMs) have shown great promise in automating unit test generation, significantly reducing the manual effort required by developers.
To effectively evaluate the capabilities of LLMs in this domain, it is crucial to have a well-designed benchmark that accurately reflects real-world scenarios and mitigates common pitfalls. 
Existing LLM test generation benchmarks are limited by two critical drawbacks: data contamination and structurally simple function code. As a result, we often cannot rely on the validity of scientific conclusions drawn from empirical studies using these limited benchmarks. 
The empirical evidence presented may be biased due to contamination and may fail to generalize beyond toy programs due to structural simplicity.

To address these problems, we introduce \dataset (UnLeakedTestbench), a new benchmark specifically designed for function-level unit test generation from real-world Python functions.
\dataset is constructed through a multi-stage curation process that ensures high cyclomatic complexity and mitigates test case contamination.
With 3,909 carefully selected function-level tasks, \dataset provides a more realistic and challenging evaluation of LLMs' test generation capabilities.
We also provide \lkdataset (PreLeakedTestbench), a pair benchmark of \dataset with leaked tests designed to enable a controlled analysis of memorization versus reasoning in test generation.
Based on the two datasets, we conduct a large-scale empirical study involving 12 state-of-the-art LLMs, comparing their performance against established benchmarks.
Our evaluation results demonstrate that \dataset is significantly more challenging. For example, test cases generated by LLMs only achieve 41.32\%, 45.10\%, 30.22\%, and 40.21\% for accuracy, statement coverage, branch coverage, and mutation score on average for all LLMs, respectively. These results are substantially lower than the corresponding metrics on TestEval (91.79\%, 92.18\%, 82.04\%, and 49.69\%) and \lkdataset (47.07\%, 55.13\%, 40.07\%, and 50.80\%). 

In addition, different from existing benchmarks, \dataset shows a strong correlation between test generation performance and code generation performance.
For example, the correlation coefficient between the coding ability and test generation performance ($Pass@1$) on \dataset is 0.79 (p = 0.002), while it is only 0.56 (p = 0.059) and 0.52 (p = 0.080) on TestEval and \lkdataset, respectively.
This indicates that \dataset more effectively measures the generalization ability of LLMs.
We also make \dataset and evaluation results publicly available to foster further research\footnote{To preserve the integrity of the benchmark and prevent test case leakage into future LLM training sets, we do not release the ground-truth tests. Instead, we provide the complete evaluation results for all models benchmarked in this paper. This allows researchers to compare new models against our findings without compromising the benchmark.}. \dataset is available at \url{https://github.com/huangd1999/UnLeakedTestBench}.
\end{abstract}

\maketitle

\section{Introduction}
\label{sec:introduction}
Reliable and robust software systems are essential in today's technology-driven world, where software failures can lead to significant financial losses, reputational damage, and even safety risks.
Effective software testing is the cornerstone of achieving this reliability, serving as a systematic approach to validate that software behaves as intended under various conditions and to identify defects before deployment \cite{anand2013orchestrated, li2017survey}.
Among the various layers of testing, unit testing holds a particularly critical position \cite{schafer2023empirical}.
Unit testing involves the creation of specific test inputs designed to scrutinize individual components or ``units'' of software, typically functions, in isolation.
The primary objective of unit test generation is to cover diverse program statements and execution branches \cite{huang2024rethinking, wang2024testeval, jain2024testgeneval}, thereby detecting defects early in the development lifecycle when they are least expensive to fix, and facilitating safer code refactoring and maintenance.
However, despite its importance, the manual composition of comprehensive unit test suites is usually labor-intensive and intellectually demanding.
Existing research and industry experience consistently highlight the significant manual effort, time investment, and domain expertise traditionally required, making it a frequent bottleneck in agile development environments \cite{wang2024testeval, daka2014survey, yu2025decon}.

To address this challenge and accelerate the unit test generation process, recent research has increasingly explored the application of LLMs in automated test case generation \cite{chen2022codet,huang2023codecot,huang2023agentcoder,du2024mercury,shinn2023reflexion,yang2025kernelgpt,xia2023universal,deng2024large,yang2024whitefox}.
With their advanced capabilities in code understanding and generation, LLMs have shown promise in automating aspects of test case generation. 
For instance, LLMs have been employed to generate syntactically and semantically valid inputs for fuzzing Deep Learning (DL) libraries by implicitly learning complex API constraints (e.g., TitanFuzz \cite{deng2023large}), and further refined to synthesize unusual programs by leveraging historical bug data (e.g., FuzzGPT \cite{deng2023fuzzgpt}).
Other applications include synthesizing syscall specifications for kernel fuzzing (e.g., KernelGPT \cite{yang2025kernelgpt}), employing multi-agent LLM frameworks for white-box compiler testing (e.g., WhiteFox \cite{yang2024whitefox}), and enhancing the general reasoning and learning capabilities of LLM agents to improve coding performance (e.g., Reflexion \cite{shinn2023reflexion}).
These diverse efforts underscore the potential of LLMs to reduce the manual effort required for test generation and enhance testing sophistication.
However, the quality of LLM-generated tests is also critical; inadequately generated tests may fail to detect crucial bugs, leading to a false sense of security and the propagation of vulnerabilities into production systems.
To effectively evaluate the capabilities of LLMs in generating unit tests, it is essential to have a well-designed benchmark that accurately reflects real-world scenarios and mitigates common pitfalls.

Several benchmarks have emerged to address this need \cite{wang2024testeval,mundler2024swt,jain2024testgeneval,zhang2024testbench}.
For example, TestEval \cite{wang2024testeval} introduced tasks based on 210 Python programs from LeetCode, focusing on achieving overall coverage, targeted line/branch coverage, and path coverage. 
TestBench \cite{zhang2024testbench} introduced a benchmark for evaluating LLMs on test generation tasks, focusing on the ability to generate tests that cover specific statements and branches in Java classes collected from real-world projects.
SWT-Bench \cite{mundler2024swt} transformed code repair tasks from SWE-Bench \cite{jimenez2023swe} into test generation tasks for issue reproduction, evaluating generated tests on whether they fail on the original code but pass on the patched version.
Similarly, TestGenEval \cite{jain2024testgeneval} adapted SWE-Bench to create tasks for full test file generation and test completion, using execution-based metrics on code from large, well-maintained repositories. 
These benchmarks have made significant contributions to the evaluation of LLMs on more realistic codebases, providing valuable insights into their capabilities and limitations.

However, despite these contributions, we argue that the conclusions of prior studies rest on a precarious scientific foundation due to several critical limitations that challenge their scientific validity.
Firstly, a foundational mismatch in evaluation granularity exists.
In software development, unit testing is typically performed at the function level, where individual functions are tested in isolation.
However, benchmarks like TestBench \cite{zhang2024testbench}, SWT-Bench \cite{mundler2024swt}, and TestGenEval \cite{jain2024testgeneval} often operate at the class, integration, or even a file level, which does not accurately reflect common practice.
Secondly, we identify the threat of insufficiently demanding evaluation (due to the ``toy'' code examples denoting relatively trivial test generation challenges).
For scientific conclusions to be valid, they must be tested on realistic code; while it is hard to determine what constitutes ``realistic'' code, many existing benchmarks are clearly unrealistic due to their low structural complexity.
Specifically, many of their functions exhibit low cyclomatic complexity, which can lead to inflated performance metrics as models are not adequately challenged to handle complex logic.

Finally, a more insidious threat is data leak-through.
A significant concern is data contamination, where the code and associated tests from public repositories may have already been part of the training corpora for many contemporary LLMs\footnote{\url{https://conf.researchr.org/details/icse-2025/icse-2025-SRC/6/Revisiting-SWE-Bench-On-the-Importance-of-Data-Quality-for-LLM-based-Code-Models}} \cite{zhou2025lessleak,ouedraogo2024large}.
This leakage compromises the reliability of an evaluation, as a model might achieve high performance due to memorization rather than genuine capability.
These combined limitations mean that previous work may be scientifically unreliable, potentially presenting a distorted view of LLM capabilities.

This paper directly and systematically confronts these issues by introducing \dataset (UnLeakedTestBench), a new benchmark that offers a more sound scientific basis on which to base future empirical evaluations of LLM-based unit test generation. 
The construction of \dataset was guided by three key principles: real-world relevance, cyclomatic complexity, and decontamination. 
Specifically, we sourced our candidate functions from The Stack v2 \cite{lozhkov2024starcoder}, a large and diverse corpus of permissively licensed source code. 
To ensure the functions in \dataset are sufficiently complex, we filtered out those with low cyclomatic complexity, retaining only those that present a meaningful challenge. 
Most importantly, to mitigate data contamination, we rigorously filtered out functions that have associated test cases already present in the training corpus. 
This process yielded a total of 3,909 function-level tasks, each with a cyclomatic complexity of at least 10, ensuring that the functions challenge an LLM's reasoning ability rather than its memorization.

To enable a controlled study of data contamination's effects, we also introduce a counterpart benchmark: \lkdataset (PreLeakedTestBench). 
This benchmark is composed of the very functions that were excluded during the decontamination phase of creating \dataset. 
These functions also meet our high-complexity criteria but are intentionally ``leaked'', as their ground-truth tests were found within the public data. 
By providing this set alongside its clean counterpart, we facilitate a direct and controlled analysis of memorization versus reasoning. 
Using \dataset and \lkdataset in tandem allows researchers to accurately measure performance inflation and achieve a more transparent and scientifically valid assessment of LLM capabilities in unit test generation.

To evaluate the effectiveness of \dataset, we conducted a large-scale empirical study involving 12 state-of-the-art LLMs, comparing their performance on \dataset against established benchmarks such as TestEval \cite{wang2024testeval}, as well as the counterpart benchmark \lkdataset.
Our experiments reveal that \dataset poses a significantly greater challenge than other benchmarks, with LLMs achieving substantially lower accuracy and code coverage. For example, test cases generated by LLMs only achieve 41.32\%, 45.10\%, 30.22\%, and 40.21\% for accuracy, statement coverage, branch coverage, and mutation score on average for all LLMs, respectively, which are substantially lower than the corresponding metrics on TestEval (91.79\%, 92.18\%, 82.04\%, and 49.69\%) and \lkdataset (47.07\%, 55.13\%, 40.07\%, and 50.80\%). 
Moreover, we observe a strong correlation between test generation performance and the code generation performance of LLMs on \dataset. For example, the correlation coefficient between the code generation performance and test generation performance ($Pass@1$) on \dataset is 0.79 (p = 0.002), while it is only 0.56 (p = 0.059) and 0.52 (p = 0.080) on TestEval and \lkdataset, respectively.
This finding underscores the effectiveness of \dataset in measuring the true test generation capabilities of LLMs, as it mitigates the influence of memorization and focuses on the model's ability to reason about and generate tests for complex, real-world code.

In summary, this paper makes the following contributions:
\begin{itemize}
    \item We present \dataset, a new, large-scale benchmark for evaluating LLM-based unit test generation, comprising 3,909 real-world Python functions. Its key innovation lies in a rigorous curation process that ensures high cyclomatic complexity and mitigates test case contamination.
    Our experiments demonstrate that \dataset poses a significantly greater challenge than existing benchmarks, revealing the current limitations of LLMs when faced with complex and uncontaminated real-world code.

    \item We also present \lkdataset, a counterpart benchmark designed to isolate the effects of data contamination while ensuring tasks are both realistic and complex. It allows for a controlled analysis of memorization versus reasoning in test generation, enabling researchers to accurately measure performance inflation and achieve a more transparent and scientifically valid assessment of LLM capabilities.

    % \item We are the first to conduct a study that directly investigates two critical threats to the scientific validity of prior work: the use of \jie{what does the following mean?}inadequately demanding code and data leak-through.
    \item We conduct a large-scale empirical study involving 12 state-of-the-art LLMs, comparing their performance on \dataset against established benchmarks such as TestEval and \lkdataset. Our results reveal that \dataset presents a significantly greater challenge, with LLMs achieving substantially lower accuracy and code coverage compared to existing benchmarks. 

    \item We provide strong empirical evidence that our design principles lead to a more reliable evaluation. We show that performance on \dataset has a strong, positive correlation with a model's intrinsic coding ability, confirming that it measures generalization. Conversely, we demonstrate that performance on contaminated data is skewed by memorization, particularly for branch coverage.
\end{itemize}

\section{Background}
In this section, we provide the necessary background to understand the context and significance of our work. We begin by discussing the importance of software testing, particularly unit testing, in ensuring software quality and reliability. We then explore the historical evolution of automated test case generation techniques, highlighting the challenges faced by traditional methods. Finally, we introduce the recent advancements in LLMs and their potential to revolutionize test case generation, while emphasizing the critical need for rigorous benchmarks to evaluate their effectiveness.

\subsection{Unit Testing in Software Engineering}
In recent years, software systems have been widely adopted across various domains, including web applications, mobile apps, embedded systems, and enterprise software. 
These systems exist in nearly every aspect of daily life and commerce, making their reliability and robustness paramount \cite{sommerville2011software}.
If the software systems fail, they can lead to significant consequences, including financial losses, reputational damage, and even risks to human safety.
To ensure software quality, software testing is the primary mechanism employed within the discipline of software engineering \cite{myers2011art}. It encompasses a systematic collection of activities designed to verify that a software system performs according to its specifications and to identify defects prior to operational deployment.
Software testing is typically structured into several distinct levels, such as unit testing, integration testing, system testing, and acceptance testing, each with a specific focus and scope, where unit testing occupies a foundational position in the development lifecycle \cite{runeson2006survey}.
The primary goal of unit testing is to validate that each unit behaves as expected under various conditions, thereby ensuring that the individual components function correctly before they are integrated into larger systems \cite{meszaros2007xunit}.
Effective unit testing can significantly reduce the number of defects that propagate to later stages of development, where they become more costly and complex to resolve \cite{myers2011art}.

\subsection{Automated Test Case Generation}
Despite unit testing can substantially improve software quality, it still poses significant challenges for software developers, i.e., the manual creation of comprehensive test suites is often labor-intensive, time-consuming, and requires a deep understanding of both the code under test and established software testing principles \cite{daka2014survey}.
The key reason is that crafting effective test cases necessitates meticulous consideration of a wide array of input values, including typical cases, boundary conditions, and potential edge cases, to ensure adequate exploration of the unit's functionality and execution paths. If the development team is under tight deadlines, they may not have sufficient time to create comprehensive test suites, leading to insufficient coverage and potentially undetected bugs. Then, the manual efforts required by the test generation become a bottleneck in many software development workflows, as developers must balance the need for thorough testing with the constraints of time and resources \cite{shamshiri2015automated}. This challenge is further compounded by the increasing complexity of modern software systems, which often involve intricate logic, numerous dependencies, and diverse input formats.

To address these challenges, the field of software engineering has long pursued the development of automated test case generation (ATG) techniques.
ATG aims to automate the test case generation process, thereby alleviating the manual burden on developers and improving the efficiency and effectiveness of software testing \cite{daka2014survey}.
Traditional ATG methods encompass a variety of methodologies, including symbolic execution \cite{king1976symbolic, chipounov2012s2e}, search-based software testing (SBST) \cite{mcminn2011search}, and fuzz testing \cite{miller1990empirical}.
Symbolic execution \cite{king1976symbolic, chipounov2012s2e} systematically explores program paths by treating input values as symbolic variables and deriving path constraints that can be solved to generate concrete test inputs. 
Search-based software testing (SBST) \cite{mcminn2011search} reformulates test generation as an optimization problem, employing metaheuristic search algorithms (e.g., genetic algorithms, simulated annealing) to discover test inputs that satisfy specific testing objectives, such as maximizing branch coverage.
Fuzz testing \cite{miller1990empirical} involves generating a large volume of random or semi-random inputs to uncover crashes, assertion violations, or unexpected program behaviors.
While these traditional ATG methods have achieved notable successes and have been incorporated into various practical development tools, they are often confronted with inherent limitations, i.e., the scalability to large and complex software systems, the difficulty in handling intricate program state or complex input constraints, and the tendency to generate tests that lack semantic relevance to real-world usage scenarios \cite{anand2013orchestrated}.
For example, symbolic execution can suffer from path explosion in programs with numerous branches and loops, rendering exhaustive exploration infeasible.
Similarly, unguided random fuzzing might be inefficient in achieving deep coverage of intricate program logic without specific domain knowledge or more sophisticated input generation strategies.

% \subsection{LLMs: A New Frontier in Test Generation}
\subsection{LLM-Driven Automated Test Generation}
To address the limitations of traditional ATG methods, recent research has increasingly turned to utilizing LLMs for automated test generation \cite{hou2024large, chen2022codet, deng2023large, huang2023agentcoder, deng2023fuzzgpt, yang2025kernelgpt, yang2024whitefox}. 
Different with traditional ATG techniques, LLMs leverage their extensive training on diverse codebases and natural language to generate test cases that are not only syntactically correct but also semantically meaningful and contextually relevant. 
Compared to traditional ATG methods, existing works in LLM-driven test generation have demonstrated significant promise and versatility across diverse testing scenarios. 
For example, TitanFuzz \cite{deng2023large} employs LLMs to generate syntactically and semantically valid inputs for fuzzing Deep Learning (DL) libraries, implicitly learning complex API constraints, while FuzzGPT \cite{deng2023fuzzgpt} refines LLMs to synthesize unusual programs for fuzzing by leveraging historical bug data. Beyond input generation, LLMs are being explored for more complex testing workflows, including the synthesis of syscall specifications for kernel fuzzing (e.g., KernelGPT \cite{yang2025kernelgpt}) and the development of multi-agent LLM frameworks for white-box compiler testing (e.g., WhiteFox \cite{yang2024whitefox}).
Another notable application includes the generation of test cases for specific functions or classes, where LLMs can produce syntactically valid inputs that cover various execution paths and edge cases. For instance, CodeT \cite{chen2022codet} and AgentCoder \cite{huang2023agentcoder} utilize LLMs to generate test cases for specific functions, while Mercury \cite{du2024mercury} employs LLMs to generate test cases for complex Python functions with intricate logic. 
These approaches have demonstrated the potential of LLMs to significantly reduce manual effort, improve the quality and relevance of generated tests, and handle complex code structures with greater intuitive facility than some traditional methods.
However, despite the promising results, the effectiveness of LLMs in test generation is not uniform and can vary significantly based on several factors, including the model architecture, training data, and specific task formulation.

\subsection{The Need for Rigorous Benchmarking}
To effectively evaluate the capabilities of LLMs in generating unit tests, it is essential to have a well-designed benchmark that accurately reflects real-world scenarios and mitigates common pitfalls. An well-constructed benchmark for LLM-based test generation should ideally provide a multi-dimensional assessment.
This includes evaluating their ability to generate correct test cases, the extent to which these tests cover the code under test (e.g., statement and branch coverage), and their effectiveness in detecting faults through metrics such as mutation testing scores \cite{wang2024testeval}. 
Recently, several benchmarks have emerged to address this need \cite{wang2024testeval,mundler2024swt,jain2024testgeneval,zhang2024testbench}. For example, TestEval \cite{wang2024testeval} introduced tasks based on 210 Python programs from LeetCode, focusing on achieving overall coverage, targeted line/branch coverage, and path coverage.
TestBench \cite{zhang2024testbench} introduced a benchmark for evaluating LLMs on test generation tasks, focusing on the ability to generate tests that cover specific statements and branches in Java classes collected from real-world projects.
SWT-Bench \cite{mundler2024swt} transformed code repair tasks from SWE-Bench \cite{jimenez2023swe} into test generation tasks for issue reproduction, evaluating generated tests on whether they fail on the original code but pass on the patched version.
Similarly, TestGenEval \cite{jain2024testgeneval} adapted SWE-Bench to create tasks for full test file generation and test completion, using execution-based metrics on code from large, well-maintained repositories.
These benchmarks have made significant contributions to the evaluation of LLMs on more realistic codebases, providing valuable insights into their capabilities and limitations.

However, despite their contributions, existing benchmarks face several critical limitations, which can undermine the reliability and generalizability of their findings.
Firstly, in software development, unit testing is typically performed at the function level, where individual functions are tested in isolation. However, benchmarks like TestBench, SWT-Bench, and TestGenEval often operate at the class, integration, or even a file level, which may not accurately reflect the common practice of unit testing at the function level.
Secondly, many of the functions in these benchmarks exhibit low cyclomatic complexity, meaning they are structurally simple and do not adequately challenge an LLM's ability to generate tests for more complex logic and diverse execution paths. This can lead to inflated performance metrics, as LLMs can achieve high coverage with relatively few, straightforward test cases.
Finally, a significant concern is data contamination, where the code and potentially associated tests from public repositories may have already been part of the training corpora for many contemporary LLMs \cite{zhou2025lessleak}. This leakage compromises the reliability of the evaluation, as LLMs might demonstrate high performance due to memorization rather than genuine test generation capabilities.
These limitations can lead to an unreliable assessment of an LLM's inherent test generation ability, as the benchmarks may not accurately reflect the challenges and complexities encountered in real-world software development.

To address these limitations and provide a more reliable benchmark, we introduce \dataset, a novel dataset constructed for evaluating LLM-driven test case generation for real-world, function-level Python tasks. \dataset is designed to mitigate the identified shortcomings of existing benchmarks by focusing on function-level testing, incorporating a diverse set of tasks with varying complexity, and ensuring that the dataset is free from data contamination. By providing a more accurate and challenging evaluation framework, \dataset aims to enhance the understanding of LLM capabilities in generating high-quality unit tests for Python code.

\section{Methodology}
\label{sec:methodology}
\subsection{Overview}
\label{sec:overview}
In this section, we present the benchmark construction process for \dataset and \lkdataset, two benchmarks designed to evaluate LLM-driven unit test generation for real-world Python functions.
Specifically, we first detail the multi-stage curation process designed to yield a high-quality dataset characterized by real-world relevance, controlled complexity, and mitigated data contamination.
Subsequently, we define the distinct task formats through which LLMs are prompted and their test generation capabilities are assessed.
Finally, we outline the specific effectiveness metrics used to quantify and compare the performance of LLMs in these tasks.

\subsection{Benchmark Construction}
\label{sec:construction}
\noindent \textbf{Data Collection} We aim to evaluate LLMs' ability to generate unit tests for real-world Python functions, focusing on their capacity to cover diverse program statements and execution branches. To achieve this, we begin by collecting candidate Python functions from The Stack v2 \cite{lozhkov2024starcoder}, a large and diverse corpus of permissively licensed source code. The Stack v2 is a vast collection of open-source code, encompassing a wide range of programming languages and domains, making it an ideal source for our benchmark. 

After collecting the candidate functions, we implement a rigorous multi-stage filtering pipeline to select suitable candidates. The filtering process is designed to ensure that the selected functions are sufficiently complex, self-contained, and free from contamination (for \dataset only), by pre-existing test cases. We introduce each step in the following.

%After applying the these filtering steps, we retained a total of 3909 function-level tasks for \dataset.

\noindent \textbf{Filtering by Cyclomatic Complexity} One of the core aims of our benchmark work is to assess an LLM's ability to generate test cases for functions with non-trivial control flow, thereby requiring tests that cover code statements and branches. Functions that are overly simplistic (e.g., possessing only a single execution path) do not offer a discerning challenge for this purpose. To address this, we follow the setup of TestEval \cite{wang2024testeval} and employ cyclomatic complexity\footnote{\url{https://en.wikipedia.org/wiki/Cyclomatic_complexity}} as a quantitative measure of a program's structural complexity. Given the control flow graph of a program, its cyclomatic complexity $V$ is defined as $V = e - n + p$, where $e$ is the number of edges, $n$ is the number of nodes, and $p$ is the number of connected components in the graph. A higher cyclomatic complexity generally indicates a greater number of decision points and potential execution paths within a function. We filter out functions with a cyclomatic complexity of less than 10. The functions retained in our dataset after all filtering stages possess an average cyclomatic complexity of 14.87, ensuring a substantive level of logical complexity for testing.

We acknowledge that the use of cyclomatic complexity can be controversial, with some studies arguing that it often serves as a proxy for code size \cite{shepperd:critique-mccabe,henderson-sellers:clarification}. 
However, we do not claim it necessarily refers to any ``conceptual complexity'' of the code. Instead, we use it strictly as a structural filter for our evaluation examples. While filtering by code size alone is an option, it would not guarantee the selection of functions with the nested Abstract Syntax Tree (AST) structures that characterize complex control flow. Without this filter, we might inadvertently include long methods that lack any significant branching substructure. By enforcing a minimum cyclomatic complexity, we explicitly select for functions that possess non-trivial branching logic. Therefore, we use it as a measure of AST complexity, and since it is also correlated with size, we ensure that as complexity increases, we are evaluating larger programs with meaningful branching. We further validate this assumption by analyzing the correlation between cyclomatic complexity and the performance of LLMs on our test generation tasks (See \cref{sec:rq2}).

\noindent \textbf{Self-Containment} For effective unit test generation, it is crucial that the function under test (FUT) can be evaluated in relative isolation, without complex external dependencies that the LLM might lack context for, such as custom functions or classes defined elsewhere in a large project. If a function heavily relies on other custom functions or classes defined elsewhere in a large project, an LLM provided only with the FUT's source code may struggle to generate correct and executable tests due to missing knowledge of these external components. Therefore, we analyzed the remaining functions to identify and filter out those that exhibited direct dependencies on or interactions with other non-standard, user-defined functions or complex class structures not passed as simple arguments. This step aims to ensure that the context provided to the LLM (primarily the function's own source code and potentially its signature) is largely sufficient for generating meaningful tests. 

\noindent \textbf{Testability Guarantee} While a function might be self-contained in terms of custom project-specific code, it often relies on standard Python libraries or widely-used third-party packages (e.g., numpy, pandas, glob, csv, logging). For each of the functions that passed the cyclomatic complexity and self-containment filters, we first verified that it was indeed testable by developing at least three distinct, executable test inputs. If a function could not be tested by the input tests, we then feed it to GPT-4o to check whether bugs exist in the function. If bugs are found, we will requires GPT-4o to fix the bugs and re-verify the testability of the function. We set the debugging limit to 3 times, i.e., if the function is still not testable after 3 times of debugging, we will discard the function. This step served as a crucial quality check, ensuring that the functions selected for the benchmark are avaliable to unit testing. During this process, we also identified and explicitly added necessary import statements for standard Python libraries and common, permissively licensed third-party libraries directly into the context provided for each function. This ensures that the code, when presented to an LLM and subsequently when its generated tests are executed, has its immediate dependencies readily available.

\noindent \textbf{Decontamination} 
The final step in our benchmark construction is the decontamination process, which allows us to define two benchmarks for a controlled analysis: \dataset and \lkdataset.
Our primary concern is mitigating the leakage of pre-existing test cases, as a model's performance could reflect memorization rather than genuine test generation aptitude if LLMs were trained on them. 
To identify contamination, we first extracted the name of each candidate function (e.g., \textit{func\_name}) and searched The Stack v2 for corresponding test definitions (e.g., \textit{def test\_func\_name}) or assertions (e.g., \textit{assert func\_name}). 
Functions for which no corresponding test cases were found form our primary benchmark, \dataset, providing a rigorous evaluation of an LLM's generalization ability. 
Our second benchmark, \lkdataset, is a superset that contains all the functions from \dataset plus all the functions that were identified as potentially contaminated. 
By including the decontaminated set within the leaked set, \lkdataset represents a benchmark with mixed data quality, allowing researchers to measure the specific impact of data contamination by comparing performance against the purely decontaminated \dataset.

After applying these steps, we retained a total of 3,909 function-level tasks for \dataset, and 18,169 function-level tasks for \lkdataset.

\subsection{Task Definition}

To measure the test generation capabilities of LLMs, we follow the setup of TestEval \cite{wang2024testeval} and define a $K$-query function-level unit test generation task. This task is designed to evaluate the model's ability to generate high-quality (i.e., correct, diverse, and effective) test cases for a given function under test (FUT).
During the evaluation, the LLM is first provided with the source code of a single function and is prompted to generate a set of test cases that cover various aspects of the function's behavior.
Then, for other queries ($2 \le i \le K$), the LLM is provided with the source code of the FUT and all previously generated test cases, and is prompted to generate a new test case that is distinct from all previously generated ones.
The detailed task definition is as follows:

\begin{itemize}
    \item \textbf{Initial Round (Round 1):}
    In the first round ($i=1$), the LLM is provided with the source code of the FUT.
    It is then prompted to generate a single unit test case for this function.

    \item \textbf{Subsequent Rounds (Round $i > 1$):}
    For each subsequent round $i$ (from $2$ to $K$), the prompt is dynamically updated and expanded.
    The new prompt contains both the original FUT and the set of all test cases that were successfully generated in the previous rounds ($1, \dots, i-1$).
    The LLM is then explicitly instructed to generate one new test case that is distinct and different from all the previously generated ones provided in the prompt's context.
\end{itemize}

This iterative process is repeated until $K$ test cases have been generated. The final output is a test suite built through a sequence of diversification requests.

\subsection{Effectiveness Metrics}
\label{sec:metrics}
We employ a suite of well-established software testing metrics to measure the quality of the LLM-generated test cases, which are designed to assess various aspects of the generated tests, including their correctness, coverage, and fault-detection capabilities.

\subsubsection{Test Generation Accuracy ($Pass@k$)}
Test generation accuracy measures the proportion of correct test cases generated by the LLM. A test is treated as ``correct'' if it compiles, runs to completion, and its assertions reflect valid expectations of the function's behavior for the given inputs.
In our benchmark, we define the accuracy metric as $Pass@k$, which represents the average number of correct tests generated per function, normalized by the total number of test cases requested ($K$).
The calculation of $Pass@k$ is based on the number of correct tests generated for each function, where $C_i$ is the number of correct tests generated for the $i$-th function (where $0 \le C_i \le K$).
If $N$ is the total number of functions in the benchmark, the accuracy is computed as follows:
\begin{equation}
\label{eq:acc}
Pass@k = \frac{\sum_{i=1}^{N} C_i}{N \times K}
\end{equation}

\subsubsection{Code Coverage ($LCov@k$ / $BCov@k$)}
In addition to accuracy, we also measure the code coverage achieved by the generated test cases.
Code coverage is an important metric in software testing, as it indicates the extent to which the generated tests exercise the code under test. We report the code coverage across all functions for two standard metrics: Line Coverage ($LCov@k$) and Branch Coverage ($BCov@k$).
\textbf{$LCov@k$} is the percentage of executable lines in the function's source code that are executed by at first $k$ queries generated test cases. 
\textbf{$BCov@k$} is the percentage of possible execution branches (e.g., from ``if'' or ``while'' statements) that are traversed by the test suite generated by the first $k$ queries.
We also provide the improvement of coverage over the first query, i.e., $\Delta_{LCov@k}$ and $\Delta_{BCov@k}$, to measure the incremental coverage achieved by each additional test case generated.

\subsubsection{Mutation Score ($Mut@k$)}
Finally, we follow the existing work \cite{jain2024testgeneval} and measure the fault-detection capability of the generated test cases using mutation score.
Mutation score is a widely used metric in software testing that evaluates the effectiveness of a test suite in detecting faults \cite{papadakis2019mutation,jia2010analysis}.
This metric involves automatically introducing small, syntactic changes (e.g., changing ``+'' to ``-'', or ``>'' to ``>='') into the function under test to create faulty versions known as ``mutants''. 
In our evaluation, we use Cosmic-Ray\footnote{\url{https://cosmic-ray.readthedocs.io/en/latest/}} to generate all possible mutants for each function in \dataset.
A test suite ``kills'' a mutant if it fails when run against the mutated code but passes on the original.
The mutation score is the percentage of non-equivalent mutants that are killed by the generated test suite. In our evaluation, for each function, we set timeout = 120 to avoid the test suite running indefinitely on a mutant.
The mutation score is calculated as follows:
\begin{equation}
\label{eq:mut}
Mut@k = \frac{\sum_{i=1}^{N} M_i}{N}
\end{equation}
where $M_i$ is the number of mutants killed by the test suite generated for the $i$-th function, and $N$ is the total number of functions in the benchmark.
A higher score indicates a more effective test suite at finding bugs.

\section{Experiment Design}
\subsection{Research Questions}
\label{sec:rq}
% Our research is driven by three core questions designed to validate \dataset by comparing it against an established benchmark across multiple dimensions.

We formulate three research questions that guide our empirical evaluation. These questions are designed to compare the performance of state-of-the-art LLMs on \dataset against established benchmarks as well as \lkdataset.

\begin{itemize}
    \item \textbf{RQ1: To what extent do LLMs struggle with test generation on \dataset compared to other benchmarks?}

    \item \textbf{RQ2: To what extent does the cyclomatic complexity of functions in \dataset influence LLM performance compared to other benchmarks?}

    \item \textbf{RQ3: How does data contamination affect the assessment of test case generation?}
\end{itemize}

\subsection{Models and Baselines}

To answer our research questions, we conduct a comprehensive study involving a diverse set of LLMs and a comparative analysis against established benchmarks.

\subsubsection{Evaluation LLMs}
We selected a wide range of 12 state-of-the-art LLMs to ensure a broad and representative evaluation.
The selection includes models of varying sizes, from smaller models with under 2 billion parameters to large models with over 30 billion parameters.
It also covers both general-purpose instruction-tuned models and models specifically specialized for code generation and understanding.
The complete list of evaluated models is detailed in \cref{tab:llms}.

\begin{table}[h!]
\centering
\caption{The set of LLMs evaluated in our study, categorized by their accessibility and originating developer.}
\label{tab:llms}
\begin{tabular}{ll}
\toprule
\textbf{Category} & \textbf{Models} \\
\midrule
% \textbf{Open-Source} & \\
\textit{CodeLlama} & \texttt{CodeLlama-7b-Instruct-hf} \\
\textit{Seed-Coder} & \texttt{Seed-Coder-8B-Instruct} \\
\textit{DeepSeekCoder} & \texttt{deepseek-coder-1.3b-instruct}, \texttt{6.7b-instruct}, \texttt{33b-instruct} \\
\textit{Gemma-3} & \texttt{gemma-3-4b-it}, \texttt{12b-it}, \texttt{27b-it} \\
\textit{Qwen2.5-Coder} & \texttt{Qwen2.5-Coder-7B-Instruct}, \texttt{14B-Instruct}, \texttt{32B-Instruct} \\
\textit{Microsoft Phi-4} & \texttt{Phi-4-mini-instruct} \\
% \midrule
% \textbf{Closed-Source} & \\
% \textit{OpenAI} & \texttt{gpt-3.5-turbo-0125}, \texttt{gpt-4o}, \texttt{gpt-4-turbo} \\
% \textit{Anthropic} & \texttt{claude-3-haiku-20240307}, \texttt{claude-3-sonnet-20240229}, \texttt{claude-3-opus-20240229} \\
\bottomrule
\end{tabular}
\end{table}

\subsubsection{Baseline Benchmarks}
% Several benchmarks \cite{wang2024testeval,jain2024testgeneval,zhang2024testbench,mundler2024swt} have been proposed to measure the capability of LLMs in generating test cases. Our evaluation focus on the Python tasks (e.g., TestBench focus on Java tasks), or not in function level unit test case generation (e.g., TestGenEval \cite{jain2024testgeneval} and SWT-Bench \cite{mundler2024swt}), and usually have lower cyclomatic complexity for the functions in the benchmark (e.g., 87.3\% of the functions in TestGenEval \cite{wang2024testeval} have cyclomatic complexity less than 10).
Several benchmarks have been proposed to measure the capability of LLMs in generating test cases, such as TestEval \cite{wang2024testeval}, TestGenEval \cite{jain2024testgeneval}, TestBench \cite{zhang2024testbench}, and SWT-Bench \cite{mundler2024swt}. 
However, these benchmarks either focus on programming languages other than Python (e.g., TestBench focuses on Java tasks) or do not specifically target function-level unit test case generation (e.g., TestGenEval and SWT-Bench). 
In this work, we then focus on the performance of the selected LLMs on \dataset against TestEval \cite{wang2024testeval}. TestEval is a well-established benchmark for evaluating LLMs on test generation tasks, focusing on Python functions from competitive programming-style problems. Similar to \dataset, the tasks in TestEval are also designed to the function-level unit test generation, making it a suitable baseline for our study. In addition, we also include \lkdataset, which is a superset of \dataset that includes all the functions from \dataset plus all the functions that were identified as potentially contaminated. This allows us to measure the specific impact of data contamination by comparing performance against the purely decontaminated \dataset.

% Additionally, many of these benchmarks contain functions with low cyclomatic complexity, which may not adequately challenge LLMs' test generation capabilities. 
% For example, 87.3\% of the functions in TestGenEval have cyclomatic complexity less than 10.

\subsection{Inference Configuration}
\label{sec:inference_config}

To ensure the fairness of our evaluation and the reproducibility of our results, we employ a consistent inference configuration across all evaluated LLMs. For generating test cases, we use a greedy decoding strategy to minimize randomness and ensure that the output is primarily a function of the model's inherent capabilities rather than stochastic sampling. Specifically, we set the decoding temperature to $0.0$. This low value encourages the model to select high-probability tokens, yielding outputs that are stable and deterministic while still allowing for some minor variation from a purely greedy approach. The maximum number of new tokens to be generated for any given task is set to $1024$, which is sufficient to generate test cases without truncation.

\begin{table}
    \centering
    \caption{RQ1.1: Accuracy of LLM-generated test cases in TestEval, \dataset, and \lkdataset. $Pass@k$ represents the percentage of the first $K$ queries generated test cases that correctly pass the function under test. The values in parentheses indicate the improvement over the $Pass@1$.}
    \label{tab:rq11}
    \resizebox{\textwidth}{!}{
    \begin{tabular}{l|rrr|rrr|rrr}
        \toprule
        \multicolumn{1}{c|}{\multirow{2}{*}{\textbf{Model\_Name}}}  & \multicolumn{3}{c|}{\textit{$Pass@1$}}                           & \multicolumn{3}{c|}{\textit{$Pass@2$}}                                & \multicolumn{3}{c}{\textit{$Pass@5$}}               \\
        \multicolumn{1}{c|}{}                                       & \textbf{\dataset} & \textbf{\lkdataset}    & \multicolumn{1}{c|}{\textbf{TestEval}} & \textbf{\dataset}    & \textbf{\lkdataset}        & \multicolumn{1}{c|}{\textbf{TestEval}} & \textbf{\dataset}  & \textbf{\lkdataset}    & \textbf{TestEval}        \\ 
        \midrule
CodeLlama-7b-Instruct-hf & 9.98 & 40.99 & 68.10 & 9.68 (-0.29) & 37.28 (-3.71)& 66.67 (-1.43) & 8.78 (-1.19) & 32.72 (-8.27)& 66.10 (-2.00) \\
Seed-Coder-8B-Instruct & 13.28 & 52.70 & 76.67 & 14.26 (+0.98) & 51.06 (-1.64)& 74.29 (-2.38) & 14.32 (+1.04) & 48.94 (-3.75)& 68.29 (-8.38) \\
gemma-3-4b-it & 11.03 & 40.52 & 30.00 & 11.33 (+0.31) & 40.18 (-0.34)& 29.05 (-0.95) & 11.78 (+0.75) & 39.35 (-1.18)& 30.48 (+0.48) \\
gemma-3-12b-it & 12.38 & 51.40 & 52.86 & 12.48 (+0.10) & 49.96 (-1.44)& 55.24 (+2.38) & 13.69 (+1.31) & 48.84 (-2.56)& 54.29 (+1.43) \\
gemma-3-27b-it & 17.83 & 53.12 & 59.05 & 20.26 (+2.43) & 53.89 (+0.78)& 60.95 (+1.90) & 21.64 (+3.81) & 54.00 (+0.89)& 60.76 (+1.71) \\
Qwen2.5-Coder-7B-Instruct & 12.48 & 52.54 & 52.38 & 13.47 (+0.98) & 51.23 (-1.32)& 57.14 (+4.76) & 14.05 (+1.57) & 49.30 (-3.25)& 53.33 (+0.95) \\
Qwen2.5-Coder-14B-Instruct & 15.32 & 57.92 & 70.00 & 15.73 (+0.41) & 56.41 (-1.51)& 61.90 (-8.10) & 15.80 (+0.48) & 54.86 (-3.07)& 55.81 (-14.19) \\
Qwen2.5-Coder-32B-Instruct & 17.83 & 57.38 & 77.62 & 15.90 (-1.93) & 55.42 (-1.96)& 66.19 (-11.43) & 16.04 (-1.79) & 53.95 (-3.44)& 59.05 (-18.57) \\
deepseek-coder-1.3b-instruct & 9.39 & 37.75 & 38.57 & 7.57 (-1.82) & 34.52 (-3.23)& 39.05 (+0.48) & 6.56 (-2.83) & 32.73 (-5.03)& 39.90 (+1.33) \\
deepseek-coder-6.7b-instruct & 10.62 & 44.84 & 55.71 & 9.38 (-1.24) & 41.23 (-3.61)& 52.62 (-3.10) & 8.60 (-2.02) & 37.48 (-7.36)& 49.81 (-5.90) \\
deepseek-coder-33b-instruct & 13.64 & 49.17 & 72.86 & 11.91 (-1.73) & 46.13 (-3.04)& 61.19 (-11.67) & 11.02 (-2.61) & 42.93 (-6.23)& 56.10 (-16.76) \\
Phi-4-mini-instruct & 8.52 & 42.76 & 39.05 & 9.21 (+0.69) & 41.18 (-1.58)& 38.57 (-0.48) & 8.55 (+0.03) & 38.33 (-4.43)& 29.24 (-9.81) \\
\midrule
Overall & 12.69 & 48.42 & 57.74 & 12.60 (-0.09) & 46.54 (-1.88) & 55.24 (-2.50) & 12.57 (-0.12) & 44.45 (-3.97)& 51.93 (-5.81) \\
        \bottomrule
    \end{tabular}
    }
\end{table}

\section{Results and Findings}
\label{sec:results}

\subsection{RQ1: To what extent do LLMs struggle with test generation on \dataset compared to other benchmarks?}\label{sec:rq1}
To answer RQ1, we compare the performance of various LLMs on \dataset against their performance on TestEval and \lkdataset\footnote{We also provide the cyclomatic complexity distribution level analysis in \cref{sec:rq2} and \cref{sec:rq3.1} to further analyze the performance of LLMs.}.

\subsubsection{RQ1.1: Accuracy of LLM-generated test cases}\label{sec:rq1.1}
We first analyze the accuracy of the test cases generated by the LLMs on \dataset, \lkdataset, and TestEval.
The evaluation results of LLM-generated test cases are shown in \cref{tab:rq11}, which demonstrate that the accuracy of LLM-generated test cases on \dataset is significantly lower than on TestEval and \lkdataset.
For example, the average $Pass@1$ across all models is 12.69\% on \dataset, compared to 48.42\% on \lkdataset and 57.74\% on TestEval. 
The $Pass@2$ and $Pass@5$ metrics show similar trends, with average scores of 12.60\% and 12.57\% on \dataset, respectively, compared to 46.54\% and 44.45\% on \lkdataset, and 55.24\% and 51.93\% on TestEval.
Next, we can also observe that the performance gap consistently exists across all model families and all metrics.
For example, for CodeLlama-7B-Instruct-hf, the $Pass@1$ score is 9.98\% on \dataset, while it is 40.99\% on \lkdataset and 68.10\% on TestEval, indicating a performance drop of 31.01\% and 58.12\% when compared to \lkdataset and TestEval, respectively.
The results indicate that LLMs struggle significantly with test generation on \dataset compared to \lkdataset and TestEval, suggesting that \dataset presents a more challenging and realistic evaluation of LLMs' test generation capabilities compared to other benchmarks.

\mybox{Answer to RQ1.1: LLMs struggle significantly with generating correct tests on \dataset compared to other benchmarks. The average $Pass@5$ score on \dataset is 12.57\%, while it is 44.45\% on \lkdataset and 51.93\% on TestEval. This suggests that \dataset presents a more challenging evaluation of LLMs' test generation capabilities.}

\begin{table}
    \centering
    \caption{RQ1.2: Code line coverage of LLM-generated test cases in three datasets. $LCov@k$ represents the percentage of executable lines in the function's source code that are executed by the first $K$ queries generated test cases. The values in parentheses indicate the improvement over the $LCov@1$.}
    \label{tab:rq_line_cov}
    \resizebox{\textwidth}{!}{
        \begin{tabular}{l|rrr|rrr|rrr}
            \toprule
            \multicolumn{1}{c|}{\multirow{2}{*}{\textbf{Model\_Name}}} & \multicolumn{3}{c|}{$LCov@1$}                   & \multicolumn{3}{c|}{$LCov@2$}                                  & \multicolumn{3}{c}{$LCov@5$} \\
            \multicolumn{1}{c|}{}                                       & \textbf{\dataset} & \textbf{\lkdataset}    & \multicolumn{1}{c|}{\textbf{TestEval}} & \textbf{\dataset}   & \textbf{\lkdataset}        & \multicolumn{1}{c|}{\textbf{TestEval}} & \textbf{\dataset} & \textbf{\lkdataset}    & \textbf{TestEval}        \\ 
            \midrule
CodeLlama-7b-Instruct-hf & 34.31 & 37.40 & 83.34 & 37.37 (+3.05) & 43.00 (+5.61) & 85.32 (+1.98) & 39.67 (+5.36) & 47.10 (+9.71) & 86.85 (+3.52) \\
Seed-Coder-8B-Instruct & 41.38 & 43.95 & 92.51 & 46.11 (+4.73) & 51.54 (+7.59) & 94.71 (+2.20) & 51.57 (+10.19) & 61.35 (+17.40) & 96.60 (+4.09) \\
gemma-3-4b-it & 35.02 & 38.66 & 71.09 & 38.90 (+3.87) & 45.55 (+6.89) & 73.35 (+2.26) & 43.92 (+8.90) & 54.74 (+16.08) & 74.78 (+3.70) \\
gemma-3-12b-it & 40.30 & 43.22 & 90.46 & 45.64 (+5.34) & 52.02 (+8.80) & 94.53 (+4.07) & 52.05 (+11.76) & 63.03 (+19.81) & 97.31 (+6.85) \\
gemma-3-27b-it & 41.09 & 44.75 & 91.13 & 46.76 (+5.67) & 53.16 (+8.40) & 94.92 (+3.80) & 53.74 (+12.65) & 64.03 (+19.28) & 97.14 (+6.01) \\
Qwen2.5-Coder-7B-Instruct & 40.73 & 43.37 & 91.47 & 45.74 (+5.01) & 51.39 (+8.03) & 94.85 (+3.38) & 51.99 (+11.27) & 62.42 (+19.05) & 97.25 (+5.78) \\
Qwen2.5-Coder-14B-Instruct & 40.64 & 43.39 & 92.35 & 45.44 (+4.80) & 51.10 (+7.71) & 94.99 (+2.63) & 52.11 (+11.47) & 62.89 (+19.50) & 97.17 (+4.82) \\
Qwen2.5-Coder-32B-Instruct & 41.32 & 44.38 & 91.79 & 47.04 (+5.72) & 53.23 (+8.85) & 95.12 (+3.32) & 52.87 (+11.55) & 63.89 (+19.51) & 97.24 (+5.45) \\
deepseek-coder-1.3b-instruct & 28.28 & 32.35 & 86.19 & 30.67 (+2.39) & 37.37 (+5.02) & 88.18 (+1.99) & 30.98 (+2.70) & 38.12 (+5.77) & 88.24 (+2.05) \\
deepseek-coder-6.7b-instruct & 25.88 & 31.72 & 90.46 & 29.05 (+3.17) & 38.48 (+6.75) & 93.12 (+2.66) & 32.83 (+6.96) & 46.86 (+15.14) & 95.10 (+4.64) \\
deepseek-coder-33b-instruct & 29.95 & 34.34 & 92.31 & 33.47 (+3.52) & 40.93 (+6.59) & 94.46 (+2.15) & 37.11 (+7.17) & 47.86 (+13.52) & 95.75 (+3.44) \\
Phi-4-mini-instruct & 35.69 & 37.86 & 80.63 & 39.02 (+3.33) & 43.54 (+5.68) & 81.62 (+1.00) & 42.35 (+6.66) & 49.32 (+11.46) & 82.69 (+2.06) \\
\midrule
Overall & 36.22 & 39.62 & 87.81 & 40.43 (+4.22) & 46.78 (+7.16) & 90.43 (+2.62) & 45.10 (+8.89) & 55.13 (+15.52) & 92.18 (+4.37) \\
            \bottomrule
        \end{tabular}
    }
\end{table}

\subsubsection{RQ1.2: Line coverage of LLM-generated test cases}\label{sec:rq1.2}
In addition to accuracy, we also evaluate the line coverage of the test cases generated by the LLMs in \dataset, \lkdataset, and TestEval. The code line coverage results of LLM-generated test cases are shown in \cref{tab:rq_line_cov}. 
We can observe that the average line coverage ($LCov@k$) of LLM-generated test cases on \dataset is lower than on TestEval and \lkdataset across all models and all $k$ values.
For example, the average $LCov@1$ across all models is 36.22\% on \dataset, while it is 39.62\% on \lkdataset and 87.81\% on TestEval, when the $K$ is set to 1. This indicates that the functions in \dataset are inherently more difficult to cover, as the LLMs struggle to generate test cases that cover a significant portion of the code lines.
In addition, we can also observe that when we increases the $K$, the $LCov@k$ of LLM-generated test cases on \dataset increases, but it remains lower than on TestEval and \lkdataset. For instance, the average $LCov@2$ on \dataset is 40.43\%, while it is 46.78\% on \lkdataset and 90.43\% on TestEval. Similarly, the average $LCov@5$ on \dataset is 45.10\%, while it is 55.13\% on \lkdataset and 92.18\% on TestEval.
This indicates that even with multiple attempts, LLMs struggle to explore the source code of the complex, real-world functions in \dataset as effectively as they do for the algorithmic problems in TestEval.

Furthermore, we can observe that the improvement in line coverage of LLM-generated test cases on \dataset is lower than \lkdataset. For example, the average improvement of line coverage on \dataset is 4.22\% and 8.89\%, while it is 7.16\% and 15.52\% on \lkdataset for $LCov@2$ and $LCov@5$, respectively. We indicate that the key reason for this is that the LLMs based on its memorization ability can generate test cases that cover more lines in \lkdataset, but they struggle to generate test cases that cover the lines in \dataset\footnote{We indicate the key reason for the lower improvement on TestEval are due to most of the lines in the function have been covered. Then, LLMs can only achieve lower improvement in line coverage for more test cases}.

\mybox{Answer to RQ1.2: LLMs struggle more with line coverage on \dataset compared to other datasets. The average $LCov@5$ on \dataset is 45.10\%, while it is 55.13\% on \lkdataset and 92.18\% on TestEval. }

\begin{table}
    \centering
    \caption{RQ1.3: Branch coverage of LLM-generated test cases in TestEval and \dataset. $BCov@k$ represents the percentage of possible execution branches (e.g., from ``if'' or ``while'' statements) that are traversed by the first $K$ queries generated test cases. The values in parentheses indicate the improvement over the $BCov@1$.}
    \label{tab:rq13}
    \resizebox{\textwidth}{!}{
        \begin{tabular}{l|rrr|rrr|rrr}
        \toprule
        \multicolumn{1}{c|}{\multirow{2}{*}{\textbf{Model\_Name}}} & \multicolumn{3}{c|}{\textit{$BCov@1$}}                 & \multicolumn{3}{c|}{\textit{$BCov@2$}}                 & \multicolumn{3}{c}{\textit{$BCov@5$}} \\
        \multicolumn{1}{c|}{}                                      & \textbf{\dataset} & \textbf{\lkdataset} & \multicolumn{1}{c|}{\textbf{TestEval}} & \textbf{\dataset}& \textbf{\lkdataset} & \multicolumn{1}{c|}{\textbf{TestEval}} & \textbf{\dataset} & \textbf{\lkdataset}  & \textbf{TestEval}   \\
        \midrule
CodeLlama-7b-Instruct-hf & 15.55 & 18.08 & 61.93 & 19.20 (+3.65) & 24.54 (+6.46) & 66.29 (+4.36) & 22.33 (+6.77) & 29.53 (+11.46) & 70.73 (+8.80) \\
Seed-Coder-8B-Instruct & 21.19 & 23.48 & 76.52 & 28.01 (+6.82) & 32.94 (+9.46) & 83.03 (+6.51) & 35.97 (+14.77) & 45.97 (+22.49) & 89.21 (+12.69) \\
gemma-3-4b-it & 15.83 & 19.04 & 48.28 & 21.21 (+5.38) & 27.37 (+8.32) & 54.07 (+5.78) & 28.52 (+12.69) & 39.64 (+20.60) & 58.77 (+10.49) \\
gemma-3-12b-it & 20.13 & 22.55 & 72.68 & 27.51 (+7.38) & 33.26 (+10.71) & 82.30 (+9.62) & 36.96 (+16.83) & 48.20 (+25.65) & 90.88 (+18.20) \\
gemma-3-27b-it & 20.82 & 23.83 & 74.97 & 28.85 (+8.03) & 34.38 (+10.54) & 83.93 (+8.96) & 39.29 (+18.47) & 49.64 (+25.81) & 90.89 (+15.92) \\
Qwen2.5-Coder-7B-Instruct & 19.60 & 22.03 & 74.18 & 26.57 (+6.97) & 31.87 (+9.84) & 82.92 (+8.74) & 35.75 (+16.15) & 46.62 (+24.59) & 90.01 (+15.83) \\
Qwen2.5-Coder-14B-Instruct & 20.31 & 22.70 & 76.82 & 27.47 (+7.16) & 32.56 (+9.87) & 83.80 (+6.98) & 37.01 (+16.70) & 48.17 (+25.47) & 90.92 (+14.10) \\
Qwen2.5-Coder-32B-Instruct & 20.40 & 23.12 & 75.47 & 28.53 (+8.13) & 34.17 (+11.05) & 84.32 (+8.85) & 37.57 (+17.17) & 49.01 (+25.89) & 91.26 (+15.79) \\
deepseek-coder-1.3b-instruct & 12.34 & 15.58 & 66.18 & 15.19 (+2.85) & 20.93 (+5.34) & 71.58 (+5.40) & 15.62 (+3.28) & 21.79 (+6.21) & 71.81 (+5.63) \\
deepseek-coder-6.7b-instruct & 12.25 & 16.07 & 73.62 & 16.63 (+4.39) & 24.08 (+8.01) & 80.62 (+7.00) & 22.16 (+9.91) & 34.97 (+18.89) & 86.79 (+13.18) \\
deepseek-coder-33b-instruct & 14.77 & 17.57 & 75.76 & 19.74 (+4.96) & 25.21 (+7.64) & 81.88 (+6.12) & 25.13 (+10.36) & 34.55 (+16.97) & 86.37 (+10.61) \\
Phi-4-mini-instruct & 16.56 & 18.28 & 60.82 & 21.37 (+4.81) & 25.30 (+7.01) & 63.66 (+2.84) & 26.30 (+9.74) & 32.80 (+14.51) & 66.82 (+6.00) \\
\midrule
Overall & 17.48 & 20.20 & 69.77 & 23.36 (+5.88) & 28.88 (+8.69) & 76.53 (+6.76) & 30.22 (+12.74) & 40.07 (+19.88) & 82.04 (+12.27) \\
        \bottomrule
        \end{tabular}
    }
\end{table}

\subsubsection{RQ1.3: Branch Coverage of LLM-generated test cases}\section{rq1.3}
Next, we analyze the branch coverage of the test cases generated by the LLMs on both \dataset and other datasets.
Compared to line coverage, branch coverage is a more complex metric that evaluates how well the test cases exercise different execution paths in the code, especially for functions with high cyclomatic complexity.
The evaluation results are shown in \cref{tab:rq13}, which reveal that the branch coverage of LLM-generated test cases on \dataset is significantly lower than on other datasets.
For example, the average branch coverage for a single generated test ($BCov@1$) across all models is 17.48\% on \dataset, compared to 20.20\% and 69.77\% on \lkdataset and TestEval, which indicates that the functions in \dataset are inherently more difficult to cover, as they often involve intricate control flows and multiple branches.
Even with multiple test cases, the branch coverage on \dataset remains lower than on \lkdataset and TestEval. For instance, the average $BCov@5$ on \dataset is 30.22\%, while it reaches 40.07\% and 82.04\% on \lkdataset and TestEval. This suggests that even with multiple attempts, LLMs struggle to explore the control flow of the complex, real-world functions in \dataset as effectively as they do for the simpler problems in other datasets.

\mybox{Answer to RQ1.3: LLMs struggle more with branch coverage on \dataset compared to other datasets. The average $BCov@5$ on \dataset is 30.22\%, while it is 40.07\% and 82.04\% on \lkdataset and TestEval, indicating that the functions in \dataset are inherently more difficult to cover due to their complex control flows.}

\subsubsection{RQ1.4: Mutation Score of LLM-generated test cases}\label{sec:rq1.4}

Finally, we evaluate the mutation score of the test cases generated by the LLMs on \dataset, \lkdataset, and TestEval\footnote{Due to the size of \lkdataset is very large, which requires substantial computational resources for mutation testing evaluation, we then randomly sample a subset (3,909 functions in \dataset + 1,000 functions from other functions in \lkdataset) for the mutation testing}.
The mutation score is a measure of how well the test cases can detect faults in the code by introducing small changes (mutants) and checking if the test cases can catch them.
As shown in \cref{tab:rq14}, the performance of LLMs on \dataset is notably weaker than on the other benchmarks. On average, the $Mut@5$ score for \dataset is only 40.21\%, significantly lower than the 50.80\% on \lkdataset and 49.69\% on TestEval. This disparity indicates that tests generated for \dataset's complex, real-world functions are less effective at identifying potential faults compared to those for the simpler or potentially memorized problems in the other datasets.

While generating more tests improves the score across all datasets, the low starting point for \dataset highlights the underlying challenge. The average $Mut@1$ score for \dataset is a mere 20.32\%, compared to 28.93\% for \lkdataset and a much higher 40.35\% for TestEval. Although this low base allows for a large absolute improvement to 40.21\% at $Mut@5$ (+19.89\%), the final score still lags considerably behind the other benchmarks. This pattern suggests that while multiple attempts can enhance test quality on \dataset, LLMs struggle to generate a single, high-quality test, and the cumulative result remains less effective than for simpler problems.

\mybox{Answer to RQ1.4: 
The tests generated by LLMs have lower mutation scores on \dataset compared to TestEval and \lkdataset. The average $Mut@5$ on \dataset is 40.21\%, while it is 50.80\% on \lkdataset and 49.69\% on TestEval. 
}

\begin{table}
    \centering
    \caption{RQ1.4: Mutation score of LLM-generated test cases in TestEval and \dataset. $Mut@k$ represents the percentage of mutants that are killed by the first $K$ queries generated test cases. The values in parentheses indicate the improvement over the $Mut@1$.}
    \label{tab:rq14}
    \resizebox{\textwidth}{!}{
        \begin{tabular}{l|rrr|rrr|rrr}
            \toprule
            \multicolumn{1}{c|}{\multirow{2}{*}{\textbf{Model Name}}} & \multicolumn{3}{c|}{\textit{$Mut@1$}} & \multicolumn{3}{c|}{\textit{$Mut@2$}} & \multicolumn{3}{c}{\textit{$Mut@5$}} \\
            \multicolumn{1}{c|}{} & \textbf{\dataset} & \textbf{\lkdataset} & \multicolumn{1}{c|}{\textbf{TestEval}} & \textbf{\dataset} & \textbf{\lkdataset} & \multicolumn{1}{c|}{\textbf{TestEval}} & \textbf{\dataset} & \textbf{\lkdataset} & \textbf{TestEval} \\
            \midrule
            Phi-4-mini-instruct & 23.84 & 28.98 & 47.18 & 29.13 (+5.29) & 37.28 (+8.30) & 51.27 (+4.09) & 36.00 (+12.16) & 47.83 (+18.85) & 54.49 (+7.31) \\
            CodeLlama-7b-Instruct-hf & 19.56 & 26.17 & 22.26 & 25.91 (+6.35) & 34.55 (+8.38) & 23.02 (+0.76) & 29.02 (+9.47) & 41.31 (+15.14) & 24.70 (+2.44) \\
            Seed-Coder-8B-Instruct & 25.66 & 26.18 & 56.71 & 36.96 (+11.30) & 39.61 (+13.43) & 64.99 (+8.28) & 52.04 (+26.39) & 57.30 (+31.12) & 71.25 (+14.54) \\
            gemma-3-4b-it & 10.56 & 18.89 & 41.80 & 13.62 (+3.06) & 27.69 (+8.80) & 46.28 (+4.48) & 20.92 (+10.36) & 41.38 (+22.50) & 49.01 (+7.21) \\
            gemma-3-12b-it & 26.24 & 24.63 & 56.71 & 38.33 (+12.09) & 38.50 (+13.87) & 64.99 (+8.28) & 52.68 (+26.45) & 57.74 (+33.11) & 71.25 (+14.54) \\
            gemma-3-27b-it & 21.21 & 27.57 & 34.18 & 28.36 (+7.14) & 37.73 (+10.15) & 36.68 (+2.50) & 39.91 (+18.69) & 55.00 (+27.43) & 40.35 (+6.17) \\
            Qwen2.5-Coder-7B-Instruct & 16.60 & 46.06 & 23.53 & 46.35 (+29.75) & 55.11 (+9.05) & 26.96 (+3.43) & 50.39 (+33.79) & 58.86 (+12.81) & 30.91 (+7.38) \\
            Qwen2.5-Coder-14B-Instruct & 22.82 & 36.36 & 41.97 & 40.70 (+17.88) & 47.08 (+10.72) & 47.31 (+5.34) & 48.84 (+26.01) & 55.81 (+19.44) & 52.72 (+10.75) \\
            Qwen2.5-Coder-32B-Instruct & 24.62 & 40.99 & 56.71 & 50.23 (+25.61) & 51.00 (+10.01) & 64.99 (+8.28) & 56.33 (+31.71) & 59.57 (+18.58) & 71.25 (+14.54) \\
            deepseek-coder-1.3b-instruct & 17.65 & 26.56 & 30.16 & 24.04 (+6.39) & 35.83 (+9.27) & 34.99 (+4.83) & 24.47 (+6.82) & 36.75 (+10.19) & 41.04 (+10.88) \\
            deepseek-coder-6.7b-instruct & 16.91 & 19.95 & 36.46 & 25.87 (+8.95) & 32.22 (+12.27) & 44.19 (+7.73) & 37.00 (+20.09) & 50.02 (+30.06) & 51.04 (+14.58) \\
            deepseek-coder-33b-instruct & 18.13 & 24.86 & 36.55 & 25.90 (+7.77) & 34.43 (+9.57) & 38.26 (+1.71) & 34.88 (+16.74) & 48.02 (+23.16) & 38.26 (+1.71) \\
            \midrule
            \textbf{Average} & 20.32 & 28.93 & 40.35 & 32.12 (+11.80) & 39.25 (+10.32) & 45.33 (+4.98) & 40.21 (+19.89) & 50.80 (+21.86) & 49.69 (+9.34) \\
            \bottomrule
        \end{tabular}
    }
\end{table}

\begin{figure*}[h]
    \centering
    \includegraphics[width=0.9\textwidth]{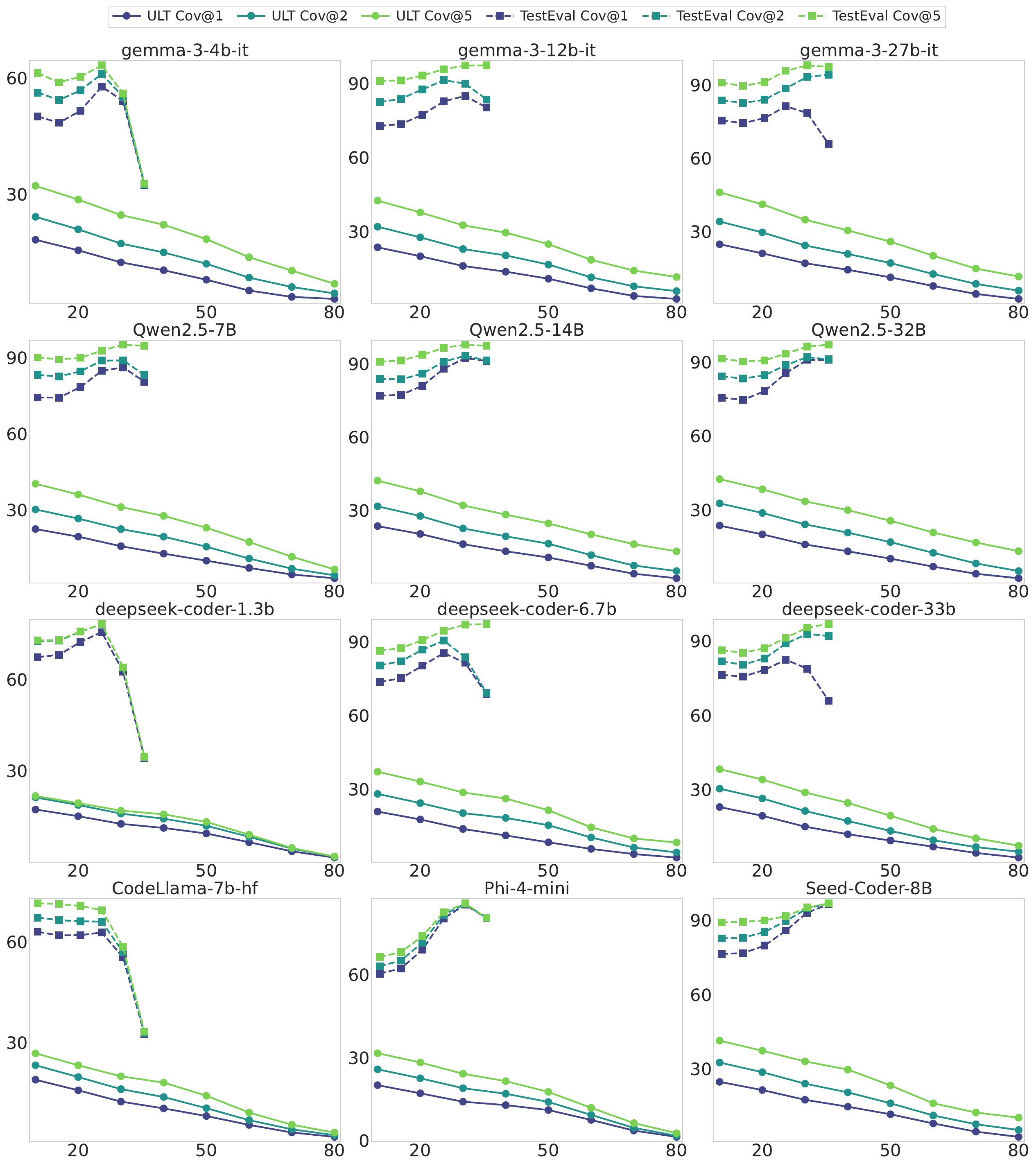}
    \caption{
        Branch Coverage as a function of Cyclomatic Complexity for \dataset and TestEval across different LLMs.
    }
    \vspace{-0.3cm}
    \label{fig:rq2_complexity}
\end{figure*}

\subsection{RQ2: To what extent does the cyclomatic complexity of functions in \dataset influence LLM performance compared to other benchmarks?}
\label{sec:rq2}

One of the key design goals of \dataset is to provide a more rigorous evaluation of LLMs by presenting them with tasks that mirror the complexity of real-world software. To investigate whether this design goal was achieved, our RQ2 examines the extent to which code complexity impacts the performance of LLM-driven test generation. We hypothesize that the higher and more diverse complexity inherent in \dataset functions poses a substantially greater challenge to LLMs compared to the more constrained, algorithmic tasks found in prior benchmarks like TestEval.
To quantify this, we use Cyclomatic Complexity, a well-established metric that measures the number of linearly independent paths through a program's source code, as a proxy for structural complexity. We then correlate this metric with the achieved Branch Coverage ($BCov@k$) of the generated test cases across various LLMs. The evaluation results are shown in \cref{fig:rq2_complexity}, which illustrates the relationship between cyclomatic complexity and branch coverage for both \dataset and TestEval across multiple LLMs\footnote{We also provide the comparison of cyclomatic complexity distributions between \dataset and TestEval in \cref{fig:rq3_complexity}.}.

As shown in \cref{fig:rq2_complexity}, the first step in our analysis is to compare the distribution of cyclomatic complexity between \dataset and TestEval.
The x-axis represents the cyclomatic complexity of the function under test, grouped into bins of 10 for \dataset and 5 for TestEval to accommodate the broader range of complexities in \dataset.
We can observe that \dataset features a significantly higher and broader complexity distribution compared to TestEval. 
For example, the cyclomatic complexity of \dataset tasks spans a wide range from 10.0 to 82.0, with a mean complexity of 14.87.
In contrast, TestEval contains tasks that are far more constrained, with its complexity concentrated in a narrow band from 10.0 to 40.0 and a lower mean of 12.35. The cyclomatic complexity of \dataset and TestEval explain the stark performance differences observed in RQ1. We also provide the cyclomatic complexity distribution of \dataset and TestGenEval in \cref{sec:discussion3}, which also shows that \dataset has a wider distribution of cyclomatic complexity than TestGenEval.

Next, we analyze the performance of LLMs on \dataset and TestEval at equivalent levels of cyclomatic complexity.
This allows us to isolate the effect of task complexity from the inherent differences in task nature between the two benchmarks.
We focus on the cyclomatic complexity range of $[10, 20)$, where both benchmarks have a sufficient number of tasks, and we can make a direct comparison.
The evaluation results are shown in \cref{fig:rq2_complexity}, where we can see a significant performance gap between the two benchmarks.
For instance, the average $BCov@5$ for CodeLlama-7B-Instruct-hf on TestEval tasks with cyclomatic complexity between 10 and 20 is close to 70\%, while the average $BCov@5$ for the same model on \dataset tasks in the same complexity range is lower than 30\%, which means that there is a about 40\% performance degradation for the same complexity level.

Furthermore, we observe a clear and consistent negative correlation between increasing complexity and test generation performance within the \dataset dataset itself.
As shown in \cref{fig:rq2_complexity}, for all models, the trend line for \dataset exhibits a steady decline as cyclomatic complexity grows.
For example, the average branch coverage ($BCov@5$) begins at 42.95\% for the simplest functions (complexity $[10, 20)$) for Qwen2.5-Coder-7B-Instruct. The performance then systematically degrades, dropping to 37.48\% for the $[20, 30)$ bin, and falling to a mere 2.40\% for highly complex functions in the $[80, 90)$ bin, which demonstrates a clear trend of performance degradation as complexity increases.
However, different with the trends observed in \dataset, the trend lines for TestEval usually exists a unusual performance behaviors. For example, for the $[10, 15)$ bin, the average $BCov@5$ for Qwen2.5-Coder-7B-Instruct is around 91.25\%, but this drops to only 87.90\% for the $[15, 20)$ bin, and then further increases to 91.67\% for the $[25, 30)$ bin.

\mybox{\textbf{Answer to RQ2:} \dataset provides a significantly higher and broader distribution of cyclomatic complexity compared to TestEval, with complexities ranging from 10.0 to 82.0 and a mean of 14.87. 
When comparing performance at equivalent complexity levels, LLMs show a substantial performance drop on \dataset, even when complexity is similar, indicating that \dataset's tasks are inherently more complex and challenging. 
For example, Qwen2.5-Coder-7B-Instruct achieves an average $BCov@5$ of 42.95\% on \dataset tasks with cyclomatic complexity between 10 and 20, while it reaches close to 70\% on TestEval.
}

\subsection{RQ3: How does data contamination affect the assessment of test case generation?}
\label{sec:rq3}

\subsubsection{RQ3.1: How does \dataset compare to \lkdataset for the same Cyclomatic Complexity?}
\label{sec:rq3.1}
\begin{figure*}
    \centering
    \includegraphics[width=0.9\textwidth]{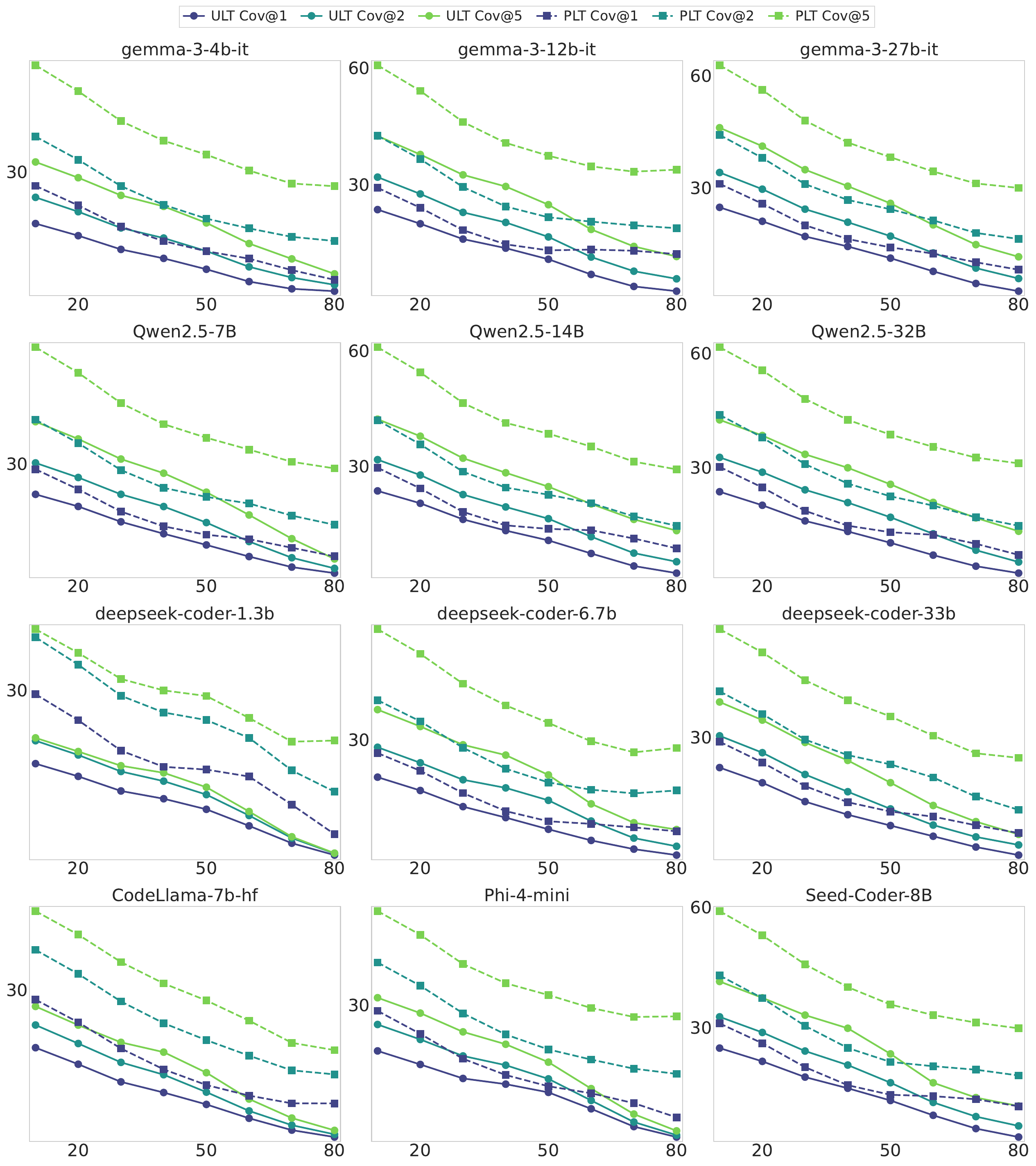}
    \caption{
        Branch Coverage as a function of Cyclomatic Complexity for \dataset and \lkdataset across four different LLMs.
    }
    \label{fig:rq3_complexity}
\end{figure*}

% One of the primary goals of \dataset is to provide a reliable evaluation of an LLM's generalization ability by mitigating data contamination.
We've compared the overall performance of \dataset and \lkdataset in RQ1. To further investigate how data contamination affects the assessment result of test case generation, we control Cyclomatic Complexity and provide further analysis.
% (**) Mark suggests: let's make "leak test bench" called "LeakedTestBench" and perhaps call unit test bench "UnLeakedTestBench". And that's encourage future authors to use the two data sets to compare and show that their techniques go beyond nearly exploiting the fact that the language model is trained on the examples they consider as a way to prove they are free from a potent flexibility. I think in this way both the leaked _and_ the unleaked data sets offer a good source of evaluation content for future empirical studies
Specifically, we provide the $BCov@k$ performance of the 12 LLMs on both \dataset and \lkdataset, as shown in \cref{fig:rq3_complexity}\footnote{To improve the readability, we focus on the data points with cyclomatic complexity in 99.5\% percentile of the cyclomatic complexity distribution of \lkdataset, which is $[10,90)$, same as the complexity range of \dataset.}.

As shown in \cref{fig:rq3_complexity}, we can observe that for all models, the branch coverage achieved by the generated test cases on \lkdataset is significantly higher than on \dataset, even when the cyclomatic complexity is equivalent.
For example, for Qwen2.5-Coder-7B-Instruct, the average $BCov@5$ on functions with cyclomatic complexity between 10 and 20 is over 55\% on \lkdataset, while it is only around 40\% on \dataset. With the increases in cyclomatic complexity, the performance gap still exists, and the branch coverage on \lkdataset is consistently higher than on \dataset.
\lkdataset and \dataset were collected in exactly the same way by the same researchers using the same process, with only one difference: whether or not the test cases were available to LLMs for training.
% mark reworded like this... I suggest to make more of this ability to measure the effect size, since so many referees get anxious, naturally so, about this leakage problem, and now you're giving all authors and reviewers a tool with which to measure the effect
Since the only difference is whether the unit tests were leaked, the most likely explanation for the performance gap is undoubtedly that the LLMs have used the leaked unit tests in pre-training.
This not only inevitably leads to artificially inflated performance metric values free LM performance on the leaked data set, it allows us to {\em empirically measure} the effect size of leakage on LLM performance.
By contrast, the lower branch coverage on \dataset suggests that the functions are more challenging and require genuine reasoning capabilities to generate effective test cases, rather than relying on memorization of previously seen test cases.
This confirms that our decontamination process was successful in creating a benchmark that genuinely challenges the models' reasoning abilities rather than their capacity for memorization.

\mybox{\textbf{Answer to RQ3.1:} \dataset has significantly lower branch coverage achieved by LLMs on \dataset compared to \lkdataset for functions of equivalent cyclomatic complexity.}

\subsubsection{RQ3.2: What is the correlation between test generation performance and code generation performance?}
\label{sec:rq3.2}
\begin{figure*}
    \centering
    \includegraphics[width=\textwidth]{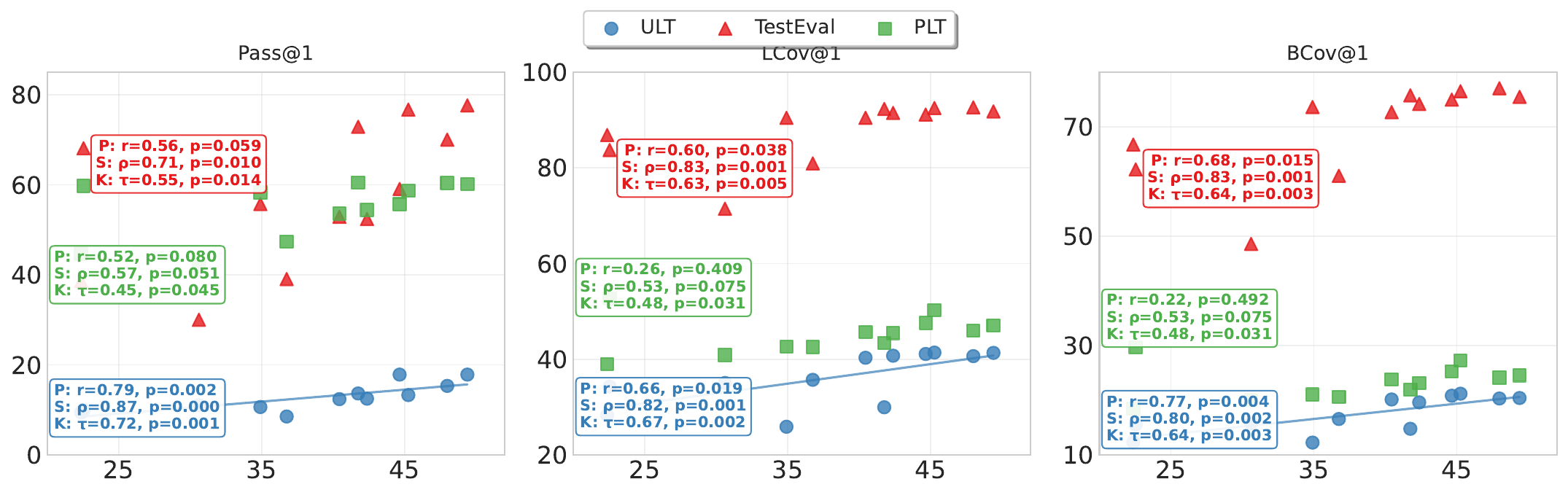}
    \caption{Correlation analysis between code generation performance on BigCodeBench (x-axis) and test generation performance metrics (y-axis). Unlike \dataset, which exhibits a strong linear correlation, there is no significant evidence for a linear correlation for TestEval nor for \lkdataset. There is some evidence for a rank correlation for TestEval and for \lkdataset, although this rank correlation is noticeably weaker than the rank correlation observed for \dataset. P means pearson correlation; K means kendall rank correlation; S means spearman correlation.}
    \label{fig:rq3_correlations}
\end{figure*}

To further investigate the effectiveness of \dataset in assessing LLMs' test generation capabilities, we analyze the correlation between test generation performance on \dataset, TestEval, and \lkdataset with LLMs' code generation performance on BigCodeBench \cite{zhuo2024bigcodebench} \footnote{We use BigCodeBench as an established proxy for an LLM's general coding ability. It is unsuitable as a test generation benchmark for our evaluation because its primary objective is to assess code generation from a prompt, often for tasks with lower structural complexity. In contrast, our focus is on a model's ability to comprehend existing, complex code to achieve high path coverage, a core challenge of unit testing that requires the high cyclomatic complexity curated in our benchmark.}.
As shown in \cref{fig:rq3_correlations} left, we first analyze the correlation between test generation accuracy ($pass@1$) and LLMs' code generation performance in BigCodeBench.
We can observe that the results reveal a clear difference between the benchmarks. For \dataset, we observe a Pearson correlation coefficient of $r = 0.79$ ($p = 0.002$), indicating a very strong and statistically significant positive correlation.

% Mark rewarded this just to make sure the statistical claims about the correlations are correct claims in case we get a statistical purist 
By contrast, the correlations for both TestEval and \lkdataset are not statistically significant, which means there is no evidence for any correlation with the available sample of data points. 
TestEval shows a correlation of $r = 0.56$ ($p = 0.059$), while \lkdataset has a correlation of $r = 0.52$ ($p = 0.080$). 
Since the number of data points in each sample is identical for these two benchmark suites, and for \dataset, we can be confident that the correlation is much stronger for \dataset and also that, if there is a correlation at all for the other two data sets, then it is considerably weaker (we would need a larger sample of data points to even precisely measure this correlation,  since it is so much weaker than that we can already confidently observe for \dataset).
The evaluation results further indicate that \dataset effectively mitigates data contamination, providing a more valid evaluation of LLMs' generalization capabilities. 
In contrast, the weaker correlations for TestEval and \lkdataset suggest that these benchmarks may not accurately reflect the models' true coding abilities, as they could be influenced by memorization of training data or other artifacts.

% Mark suggetss: include spearman and kendal rank correlatio analysis for completeness: you may find that the results are slightly more nuanced since they're probably a rank correlation even though there isn't a linear one the observable (with 95% confidence)  with this number of data points

Next, we extend our analysis to code coverage metrics, which provide a deeper understanding of the quality of generated tests beyond mere correctness. For line coverage ($lcov@1$), we can observe that \dataset exhibits a moderately strong and significant correlation with code generation performance ($r = 0.66$, $p = 0.019$).
However, for \lkdataset, the correlation is weak and not statistically significant ($r = 0.26$, $p = 0.409$), which indicates that LLMs try to memorize existing test cases from their training data to achieve high line coverage on these functions.
Similar to the line coverage, we also analyze the correlation of branch coverage ($bcov@1$) with code generation ability.
As shown in \cref{fig:rq3_correlations}, \dataset maintains a strong positive correlation with coding ability ($r = 0.77$, $p = 0.004$), which shows that for unseen and complex functions, stronger models are significantly better at navigating intricate logic to cover more branches.
For the \lkdataset, the correlation is almost negligible ($r = 0.22$, $p = 0.492$), indicating that LLMs can achieve high branch coverage on these functions without necessarily understanding the underlying logic, likely due to the memorization of existing test cases from their training data.

\mybox{\textbf{Answer to RQ3.2:}
The correlation between test generation accuracy on \dataset and LLMs' coding ability is strong ($R=0.79$ and $p=0.002$), while TestEval and \lkdataset show weaker correlations ($R=0.56$ and $R=0.52$, respectively), which indicates that \dataset effectively measures generalization capabilities rather than memorization.
In contrast, TestEval and \lkdataset exhibit weaker correlations, suggesting that they may not accurately reflect the models' true test generation abilities due to data contamination.
}

\section{Discussion}

\subsection{Impact of Iterative Query Count on Benchmark Suitability}
\label{sec:discussion1}
\begin{figure*}[!t]
    \centering
    \includegraphics[width=\textwidth]{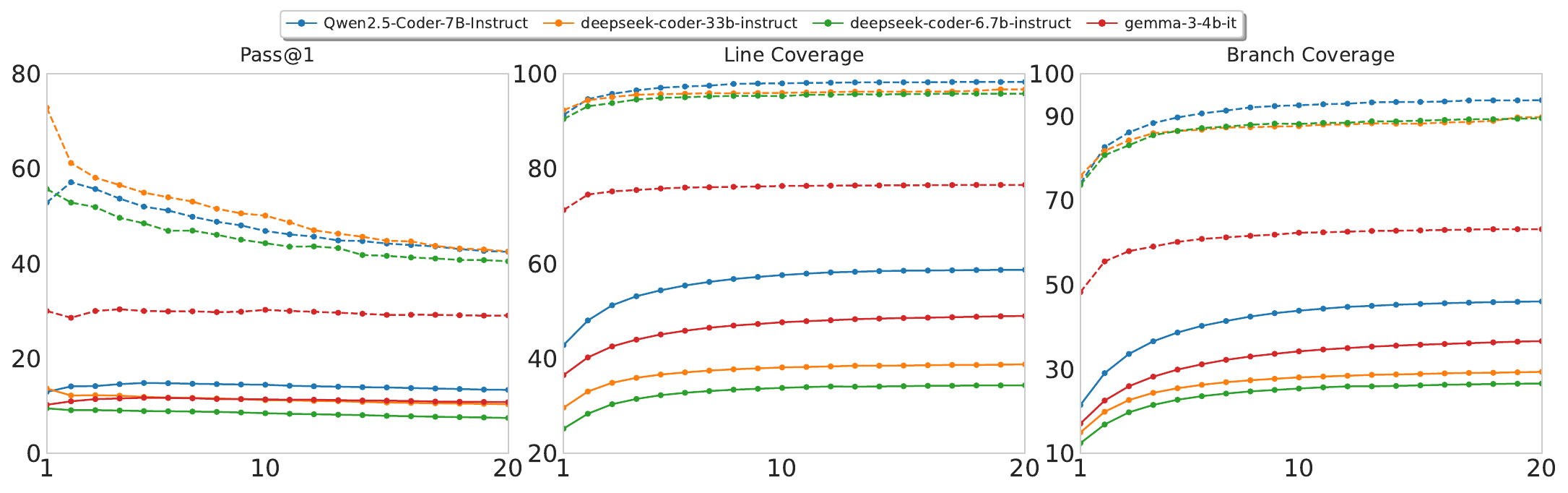}
    \caption{
        Performance trends on \dataset (solid lines) and TestEval (dashed lines) as the number of generated test cases ($k$) increases from 1 to 20. As the overhead of mutation testing for 20 times execution is very large, we do not conduct the mutation testing for this evaluation.} 
    \label{fig:discussion1_trends}
\end{figure*}

To understand how the number of iterative queries ($k$) affects the performance of LLMs on \dataset and TestEval, we conducted a detailed analysis of performance trends as $k$ was varied from 1 to 20. The evaluation results, shown in \cref{fig:discussion1_trends}, highlight two critical and divergent patterns between \dataset and TestEval.

First, as shown in \cref{fig:discussion1_trends}, we observe a significant difference in how code coverage evolves on the two benchmarks.
On TestEval, both line and branch coverage exhibit rapid initial growth and then quickly converge towards a plateau.
For instance, powerful models like Qwen2.5-Coder-7B-Instruct achieve nearly 95\% line coverage after generating just 8-10 test cases, with subsequent tests yielding diminishing returns.
This saturation indicates that the underlying functions in TestEval are structurally simple, possessing a limited number of execution paths that are easily exhausted by a small set of tests.
Consequently, TestEval has a low ceiling for evaluation, making it less suitable for assessing an LLM's ability to generate a truly comprehensive and diverse test suite over extended interactions.
In contrast, code coverage on \dataset demonstrates a sustained growth trajectory as $k$ increases.
Even after generating 20 distinct test cases, the total line and branch coverage for all models remains below 60\%.
This pattern strongly suggests that the functions within \dataset are significantly more complex, containing a large number of independent execution paths that require continuous generation of novel test inputs to be explored.
The lack of a coverage plateau provides a much longer and more discerning runway for evaluation.
It allows for a clearer differentiation between models based on their ability to consistently reason about complex code and diversify their test generation strategies over time.
This characteristic validates \dataset as a more challenging and appropriate benchmark for rigorously measuring the test generation capabilities of advanced LLMs.

The second observation is the divergent trend in test accuracy ($Pass@k$) between the two benchmarks, as shown in \cref{fig:discussion1_trends}.
On \dataset, the accuracy of generated tests remains relatively low but stable across all values of $k$.
This behavior is consistent with models confronting a genuinely difficult and unseen task; their ability to produce a correct test does not significantly degrade as they are asked for more examples.
Conversely, on TestEval, we observe a sharp and consistent decay in accuracy as $k$ increases.
For example, the accuracy for deepseek-coder-33b-instruct plummets from an initial 72.86\% at $k=1$ to 42.55\% at $k=20$.
We hypothesize that this accuracy degradation is a direct symptom of data contamination within the TestEval benchmark.
In the initial query ($k=1$), LLMs are likely to retrieve and output a memorized solution—the ``leaked'' test case they have seen during pre-training.
At this stage, the evaluation is primarily measuring the model's memorization ability, resulting in an inflated accuracy score.
However, as the iterative process continues, the prompt explicitly requires the LLM to generate a test case that is \textit{different} from the ones it has already provided.
This constraint forces the model to move beyond simple recall and engage in genuine test case generation.
At this point, the task transitions from a test of memory to a true test of reasoning and generalization.
The subsequent, significant drop in accuracy reflects the model's actual, and much lower, capability for this more complex task.
This phenomenon underscores the critical importance of our decontamination process and validates that \dataset provides a more faithful and realistic assessment of an LLM's test generation skills.

\subsection{Query Strategy}
\label{sec:discussion2}
In our experiments, to ensure that LLMs generate diverse test cases that can achieve high coverage, we requires LLMs to generate multiple test cases iteratively, with the instruction to produce a new test case that is ``different'' from the previous ones.
To explore the effectiveness of this strategy, we compare it against a more simpler approach where the model is simply asked to generate a test case for each query without any additional guidance. To avoid LLM repeat its previous test cases, we adjust the temperature to 0.2, which encourages the model to produce diverse outputs while still allowing it to generate valid test cases.
The evaluation results are shown in \cref{tab:discussion_feedback}, we can observe that our feedback-driven query strategy (w/ Feedback) significantly outperforms the way where LLMs are simply asked to generate different test cases (w/o Feedback) across all metrics and models.
For example, for Qwen2.5-Coder-7B-Instruct, the line coverage ($LCov@5$) improved from 16.65\% to 21.32\% with feedback, a gain of +4.67\%, while the improvement of line coverage without feedback was only +1.08\%.
Similarly, the branch coverage ($BCov@5$) increased from 9.05\% to 15.91\% with feedback, a gain of +6.86\%, while the improvement without feedback was only +1.62\%.
This trend is consistent across both models and all metrics, indicating that the feedback-driven query strategy is significantly more effective in guiding LLMs to generate high-quality test cases that cover more lines and branches of the code under test.

\begin{table*}
\centering
\caption{Comparison of LLM performance with and without iterative feedback. The results are shown for two models, Qwen2.5-Coder-7B-Instruct and deepseek-coder-6.7b-instruct, under two different strategies: ``w/o Feedback'' and ``w/ Feedback''. The ``w/o Feedback'' strategy uses the first query prompt asking for a different test case, while the ``w/ Feedback'' strategy incorporates feedback from the previous query generated test cases.}
\label{tab:discussion_feedback}
\resizebox{\textwidth}{!}{
\begin{tabular}{l|l|rrr|rrr|rrr}
\toprule
\multicolumn{1}{c|}{\multirow{2}{*}{\textbf{Model Name}}} & \multicolumn{1}{c|}{\multirow{2}{*}{\textbf{Strategy}}} & \multicolumn{3}{c|}{\textbf{Test Accuracy ($Pass@k$)}} & \multicolumn{3}{c|}{\textbf{Line Coverage ($LCov@k$)}} & \multicolumn{3}{c}{\textbf{Branch Coverage ($BCov@k$)}} \\
\multicolumn{1}{c|}{} & \multicolumn{1}{c|}{} & \textbf{$k=1$} & \textbf{$k=2$} & \textbf{$k=5$} & \textbf{$k=1$} & \textbf{$k=2$} & \textbf{$k=5$} & \textbf{$k=1$} & \textbf{$k=2$} & \textbf{$k=5$} \\
\midrule
\multirow{2}{*}{Qwen2.5-Coder-7B-Instruct} & w/o Feedback & 5.45 & 5.26 & 5.08 & 17.04 & 17.58 \textcolor{blue}{(+0.54)} & 18.12 \textcolor{blue}{(+1.08)} & 9.45 & 10.20 \textcolor{blue}{(+0.75)} & 11.07 \textcolor{blue}{(+1.62)} \\
& w/ Feedback & 5.81 & 6.06 & 6.42 & 16.65 & 18.77 \textcolor{red}{(+2.12)} & 21.32 \textcolor{red}{(+4.67)} & 9.05 & 12.00 \textcolor{red}{(+2.95)} & 15.91 \textcolor{red}{(+6.86)} \\
\midrule
\multirow{2}{*}{deepseek-coder-6.7b-instruct} & w/o Feedback & 4.14 & 3.39 & 2.63 & 9.93 & 10.31 \textcolor{blue}{(+0.38)} & 10.57 \textcolor{blue}{(+0.64)} & 5.17 & 5.70 \textcolor{blue}{(+0.53)} & 6.06 \textcolor{blue}{(+0.89)} \\
& w/ Feedback & 4.02 & 3.82 & 3.73 & 9.94 & 11.40 \textcolor{red}{(+1.46)} & 12.96 \textcolor{red}{(+3.02)} & 5.23 & 7.22 \textcolor{red}{(+1.99)} & 9.52 \textcolor{red}{(+4.29)} \\
\bottomrule
\end{tabular}
}
\end{table*}

\subsection{Comparison with TestGenEval}
\label{sec:discussion3}
To further validate the complexity and realism of \dataset, we compared its cyclomatic complexity distribution with that of TestGenEval \cite{jain2024testgeneval}, another recent benchmark for test generation.
As shown in \cref{tab:complexity_comparison}, we can observe that \dataset has a significantly higher cyclomatic complexity range, with a mean complexity of 14.87 and a maximum of 82, compared to TestGenEval's mean complexity of 4.71 and maximum of 35.
This indicates that \dataset contains more complex functions that require advanced reasoning capabilities to generate effective test cases.
Furthermore, the cyclomatic complexity of \dataset is concentrated in a narrower range, with 100\% of its functions having a cyclomatic complexity of 10 or higher, while TestGenEval has 87.3\% of its functions with a cyclomatic complexity of 9 or less.
This further supports our claim that \dataset provides a more challenging and realistic benchmark for function-level unit test generation, as it requires LLMs to navigate complex logic and cover multiple execution paths within individual functions, rather than simply generating tests for larger files with lower complexity.

\begin{table*}
\centering
\caption{Comparison of cyclomatic complexity distribution between \dataset and TestGenEval.}
\label{tab:complexity_comparison}
% \resizebox{\textwidth}{!}{
\begin{tabular}{l|rrrrrrr}
\toprule
Benchmark & Range & < 10 & >=10 & Mean & Median & Min & Max \\
\midrule
\dataset & 10-82 & 0 & 100 & 14.87 & 12 & 10 & 82 \\
TestGenEval & 0-35 & 87.3& 12.7 & 4.71 & 3 & 1 & 35 \\
\bottomrule
\end{tabular}
% }
\end{table*}

\subsection{Avoiding Future Data Contamination}
A critical challenge for any contemporary benchmark is ensuring its long-term viability in an era where public data is continuously scraped for training next-generation LLMs. 
This process of data contamination or leakage, where a model is inadvertently trained on the benchmark it is meant to be evaluated against, can invalidate results and obscure true scientific progress. 
We have designed the release and evaluation process for \dataset specifically to mitigate this threat.

Our primary strategy is to separate the benchmark's problems from its solutions. 
We publicly release the curated set of 3,909 Python functions that serve as the evaluation problems. 
However, we deliberately withhold the ground-truth test suites that we created for our internal validation. 
These reference tests are not required for evaluation. 
Instead, we provide a self-contained evaluation script. 
Researchers use this script to evaluate the tests generated by their models; the script dynamically executes the generated tests against the functions-under-test and calculates metrics, such as line and branch coverage, without ever exposing a static set of correct solutions that could be crawled and memorized by future models. 
This approach ensures that \dataset remains a robust test of generation capability, not memorization.

This design also addresses the secondary threat of adversarial contamination, where a user might generate their own high-quality tests and add them to public corpora. 
While this is impossible to prevent entirely, our approach makes such contamination significantly more difficult and less direct than if we had provided a canonical golden set of tests ourselves. 
Furthermore, to foster a community-wide commitment to the benchmark's integrity, the license for \dataset will explicitly discourage users from making generated test suites publicly available in a manner that facilitates web crawling. 
We believe establishing this community norm is crucial for preserving the benchmark's value, balancing the principles of open science and replicability with the practical need for contamination-resistant evaluation.

\subsection{Threats to Validity}
\label{sec:threats}

\subsubsection{Internal Validity}
Internal validity concerns potential confounding factors within our experimental setup that could influence the observed outcomes.
A primary threat pertains to the reproducibility and determinism of the test cases generated by LLMs.
The inherently stochastic nature of some decoding strategies could lead to variability in results, making it difficult to attribute performance differences solely to model capabilities.
To mitigate this, we employed a greedy decoding strategy by setting the temperature parameter to 0 for all experiments, as detailed in \cref{sec:inference_config}.
This approach significantly minimizes randomness, ensuring that the generated outputs are as deterministic as possible and primarily reflect the model's core reasoning abilities rather than sampling artifacts. Another potential threat is the reliability of our test execution and evaluation environment.
Inconsistencies in the environment, such as differing library versions or system configurations, could lead to spurious test failures or inaccurate coverage measurements.
To address this, all generated test cases were executed within a unified and standardized Docker environment.
This containerized setup guarantees that every test is run against the exact same version of the Python interpreter and dependent libraries, thereby ensuring the consistency and comparability of our results across all models and benchmarks.

\subsubsection{External Validity}
External validity relates to the generalizability of our findings beyond the specific context of this study.
One threat is the representativeness of our benchmark.
Although \dataset is constructed from a large corpus of real-world Python functions from The Stack v2 and curated based on specific criteria like cyclomatic complexity and test case decontamination, it may not encompass the full spectrum of programming paradigms, application domains, or coding styles found in all software projects.
Therefore, while our results provide strong evidence regarding complex, self-contained functions, caution should be exercised when generalizing them to other types of software, such as large-scale enterprise systems or highly specialized domains.
Another threat to external validity is our focus on a single programming language, Python.
The capabilities of LLMs in code understanding and generation can vary across different languages due to differences in syntax, semantics, and representation in their training data.
The findings and performance gaps observed in this study for Python may not directly translate to other languages like Java, C++, or JavaScript.
Finally, our evaluation is limited to the specific set of 12 LLMs listed in \cref{tab:llms}.
The field of LLMs is evolving at an exceptionally rapid pace, with new and more powerful models being released frequently.
While our selection represents a broad and diverse snapshot of the current state-of-the-art, our specific findings may not be generalizable to future, more advanced models that may overcome some of the challenges identified in this work.

\subsubsection{Construct Validity}
Construct validity examines whether our evaluation metrics and experimental design accurately measure the concepts they purport to assess, namely test generation quality and a model's reasoning capability.
A potential threat is our reliance on code coverage (line and branch) and mutation score as primary proxies for the quality of a generated test suite.
While these are standard and widely accepted metrics in software testing research, they are not perfect.
High coverage does not guarantee the absence of bugs, and a test suite could kill many mutants without necessarily reflecting all aspects of real-world fault-finding effectiveness.
Nonetheless, these metrics provide a quantitative, objective, and reproducible basis for comparison that is standard practice in the field.
Another threat lies in our use of performance on BigCodeBench as a proxy for an LLM's intrinsic, general-purpose coding ability in our RQ3 analysis.
While BigCodeBench is a comprehensive benchmark, a model's proficiency in general code generation may not perfectly correlate with its specialized ability to perform test generation, which requires different reasoning skills (e.g., identifying edge cases and defining assertions).
However, by using a well-established, independent benchmark, we establish a reasonable baseline for a model's fundamental reasoning capabilities, allowing us to effectively test our hypothesis regarding data contamination.
Finally, the design of our iterative test generation task, where we prompt for a ``new'' and ``distinct'' test case in each round, could be a threat.
The interpretation of these terms by the LLM might vary, and the prompt structure itself could influence the diversification strategy of the models.
We acknowledge this as an inherent aspect of using natural language to prompt LLMs and have kept the prompt consistent across all models to ensure a fair comparison.

\section{Related Work}
\label{sec:related_work}
Our research is situated within the rapidly growing field of applying LLMs to the software engineering task of automated test generation.
This area has seen a surge of innovation, with researchers exploring various strategies to leverage the code and reasoning capabilities of LLMs.
This section reviews the literature by first surveying the landscape of LLM-based test generation techniques and then discussing the evolution of benchmarks used to evaluate them, thereby positioning the unique contribution of \dataset.

\subsection{LLMs for Test Generation}
The application of LLMs to generate unit tests has evolved from initial feasibility studies to a variety of sophisticated techniques aimed at improving the quality and effectiveness of the generated test suites.
These approaches can be broadly categorized by whether they focus on refining the model's input (prompt engineering), enhancing the model's output (post-processing and repair), or altering the generation process itself.
A foundational stream of research has focused on empirically evaluating the baseline performance of LLMs and building practical tools.
Studies by Siddiq et al. \cite{siddiq2024using}, Schäfer et al. \cite{schafer2023empirical}, and El Haji et al. \cite{el2024using} provided crucial early insights into the capabilities and limitations of models like Codex, GPT-3.5, and GitHub Copilot for generating tests in languages such as Java, JavaScript, and Python.
This foundational work paved the way for IDE plugins like TestSpark \cite{sapozhnikov2024testspark}, which integrate test generation directly into the developer's workflow.
Other work has focused on optimizing the input to the LLM to elicit higher-quality outputs.
Researchers have shown that careful prompt engineering, such as providing contextual information about the code or its dependencies, can significantly improve the quality of generated tests \cite{bareiss2022code,huang2024measuring}.
More advanced techniques employ retrieval-augmented generation, where relevant few-shot examples are dynamically retrieved from a corpus and included in the prompt to guide the model, as demonstrated by CEDAR \cite{nashid2023retrieval}.
Recognizing that LLMs often produce imperfect or incomplete tests, another major line of work focuses on enhancing or repairing the generated output.
This includes hybrid approaches that combine LLMs with traditional software testing techniques.
For instance, Codamosa \cite{lemieux2023codamosa} uses LLMs to generate new test inputs to help Search-Based Software Testing (SBST) escape from local optima and improve coverage.
Other techniques focus on iterative refinement, where the LLM's output is executed, and the resulting feedback (e.g., compilation errors, failed assertions) is used to prompt the model for a revised solution, a strategy employed by Testart \cite{gu2024testart} in a co-evolutionary cycle.
Guidance from external analysis is also common, with researchers using static analysis \cite{pan2025aster} or feedback from mutation testing \cite{dakhel2024effective} to guide the LLM toward producing more effective and bug-finding tests.
Although these techniques have shown promising results in test generation, there still lack a comprehensive benchmark that can effectively evaluate the true reasoning and generalization capabilities of LLMs in this context.
Our work addresses this gap by introducing \dataset, a benchmark specifically designed to challenge LLMs with complex, real-world functions while mitigating the risks of data contamination and ensuring a more accurate assessment of their test generation abilities.

\subsection{Benchmarks for Test Generation}
Concurrent with the development of new generation techniques has been the creation of benchmarks to evaluate their effectiveness.
An early and influential benchmark, TestEval \cite{wang2024testeval}, established a foundation by proposing tasks based on competitive programming problems from LeetCode.
It introduced key evaluation metrics, including coverage-oriented tasks and the use of mutation scores, providing a standardized basis for comparing LLMs.
However, as we demonstrated in our study, its reliance on algorithmic problems limits its representativeness of real-world software engineering challenges.
Subsequent benchmarks sought to improve real-world relevance by sourcing tasks from large, open-source software projects.
SWT-Bench \cite{mundler2024swt} and TestGenEval \cite{jain2024testgeneval} both derive their tasks from the SWE-Bench dataset \cite{jimenez2023swe}.
SWT-Bench frames the task as issue reproduction, where the goal is to generate a test that fails on buggy code but passes on the fixed version.
TestGenEval focuses on test file generation and completion, using execution-based metrics on code from major software repositories.
While these benchmarks marked a significant step towards realism, they introduced other confounding factors that \dataset is explicitly designed to address.
As discussed in our introduction and supported by our findings in RQ3, benchmarks derived from popular public repositories are susceptible to test case contamination, where models may score well by recalling solutions from their training data rather than by demonstrating genuine reasoning.
Furthermore, their file- or repository-level task granularity often leads to excessively long input contexts, which can degrade LLM performance, and their tasks may not consistently feature high structural complexity.
\dataset differentiates itself from this prior work by simultaneously ensuring real-world relevance, mitigating test case contamination through a rigorous filtering process, controlling for complexity by filtering on cyclomatic complexity, and utilizing a function-level focus to enable a more precise assessment of an LLM's core test generation capabilities.

\section{Conclusion}
\label{sec:conclusion}
In this paper, we introduce \dataset, a new benchmark specifically designed for function-level unit test generation from real-world Python functions.
\dataset is designed to address critical limitations in existing benchmarks, such as test case data contamination and insufficient program complexity.
By focusing on real-world functions with high cyclomatic complexity, \dataset provides a more challenging and realistic evaluation environment for LLMs.
We also provide \lkdataset, a pair benchmarks of \dataset with leaked tests designed to enable a controlled analysis of memorization versus reasoning in test generation.
We conducted a comprehensive experimental evaluation involving 12 state-of-the-art LLMs, systematically comparing their performance on \dataset against other benchmarks like TestEval, BigCodeBench, and \lkdataset.
Our findings reveal that \dataset significantly outperforms existing benchmarks in terms of complexity and challenge, with a broader distribution of cyclomatic complexity ranging from 10.0 to 82.0 and a mean of 14.87, compared to TestEval's concentration between 9.0 to 45.0, with a mean of 12.35.
We also demonstrated that performance on \dataset is substantially lower across all metrics (accuracy, line coverage, and branch coverage) compared to TestEval and \lkdataset, confirming that its tasks are inherently more difficult.
For example, test cases generated by LLMs only achieve 41.32\%, 45.10\%, 30.22\%, and 40.21\% for accuracy, statement coverage, branch coverage, and mutation score on average for all LLMs, respectively. These results are substantially lower than the corresponding metrics on TestEval (91.79\%, 92.18\%, 82.04\%, and 49.69\%) and \lkdataset (47.07\%, 55.13\%, 40.07\%, and 50.80\%). 
This performance gap persists even when comparing tasks of similar cyclomatic complexity, indicating that \dataset's tasks are not only more complex but also require deeper reasoning and generalization capabilities from the models.

% \balance
\bibliographystyle{IEEEtran}
\bibliography{acmart-primary/acmart}

\end{document}